\documentclass[10pt,journal,compsoc]{IEEEtran}
\usepackage{caption}
\usepackage{subcaption}

\ifCLASSOPTIONcompsoc
  \usepackage[nocompress]{cite}
\else
  \usepackage{cite}
\fi
\ifCLASSINFOpdf
    \usepackage[pdftex]{graphicx}
    \graphicspath{{images/}}
\else
\fi
\begin{document}
%
\title{Communicating Uncertainty and Risk\\ in Air Quality Maps}
%
%
%
%

\author{Annie Preston
        and~Kwan-Liu Ma}
\IEEEtitleabstractindextext{%
\begin{abstract}Environmental sensors provide crucial data for understanding our surroundings. For example, air quality maps based on sensor readings help users make decisions to mitigate the effects of pollution on their health. Standard maps show readings from individual sensors or colored contours indicating estimated pollution levels. However, showing a single estimate may conceal uncertainty and lead to underestimation of risk, while showing sensor data yields varied interpretations. We present several visualizations of uncertainty in air quality maps, including a frequency-framing ``dotmap'' and small multiples, and we compare them with standard contour and sensor-based maps. In a user study, we find that including uncertainty in maps has a significant effect on how much users would choose to reduce physical activity, and that people make more cautious decisions when using uncertainty-aware maps. Additionally, we analyze think-aloud transcriptions from the experiment to understand more about how the representation of uncertainty influences people's decision-making. Our results suggest ways to design maps of sensor data that can encourage certain types of reasoning, yield more consistent responses, and convey risk better than standard maps.
\end{abstract}

}

\maketitle

\IEEEdisplaynontitleabstractindextext

%
\IEEEpeerreviewmaketitle

\IEEEraisesectionheading{\section{Introduction}\label{sec:introduction}}

%
%
%
%
Air pollution is a ``slow disaster,'' affecting more people than widely understood~\cite{learning_from_disaster}. Worldwide, $90\%$ of people breathe polluted air, contributing to an annual death toll of 7 million~\cite{who_report}. The particulate matter PM2.5 is an especially insidious pollutant causing long-term health problems~\cite{who_report_pm25}. 
Though governments have made some successful efforts to reduce pollution, air quality in many countries is getting worse, and recent research has highlighted previously underestimated health risks and inequities from air pollution~\cite{Powell201}~\cite{Tessum6001}. Despite progress reducing emissions from cars in the United States through 2016, traffic-related pollution contributed to 
one-fifth of childhood asthma cases nationwide~\cite{asthma_2016}. Small increases in long-term exposure to PM2.5 may lead to a large increase in the COVID-19 death rate~\cite{Wu2020.04.05.20054502}.



Physicians have advocated for better tools to inform their patients about air pollution and its dangers~\cite{Powell201}. Informatics and mapping are crucial for communicating environmental hazards like air pollution, yet information is often presented in a way that reinforces biases, including underdisclosing risk~\cite{Soden:2018:CIA:3170427.3173027}. As with all disasters, air pollution ``reflects the underlying stratification of a society,'' with marginalized groups most at risk~\cite{learning_from_disaster}. Equipped with better maps, people could make more informed choices about limiting exposure, understand sources and characteristics of pollution, and reduce their own contributions to poor air quality.

In this study, we explore whether including uncertainty in maps of air quality---and potentially in maps of other sensor data---could help address the need to better communicate risk. Our contributions include:
\begin{itemize}
\item A mixed methods user study with visualization designs that vary the amount of uncertainty shown;
\item A quantitative analysis of decision making with geospatial uncertainty visualizations;
\item A qualitative think-aloud analysis illuminating how people make decisions about uncertain maps;
\item Evidence about ways to visualize this uncertainty and elicit certain decision-making patterns.
\end{itemize}

\section{Background}
\subsection{Air Quality Communication}
\label{aq_vis_and_uncert}
\begin{figure}[b]
\centering
    \begin{subfigure}{0.4\columnwidth}
        \includegraphics[width=\columnwidth]{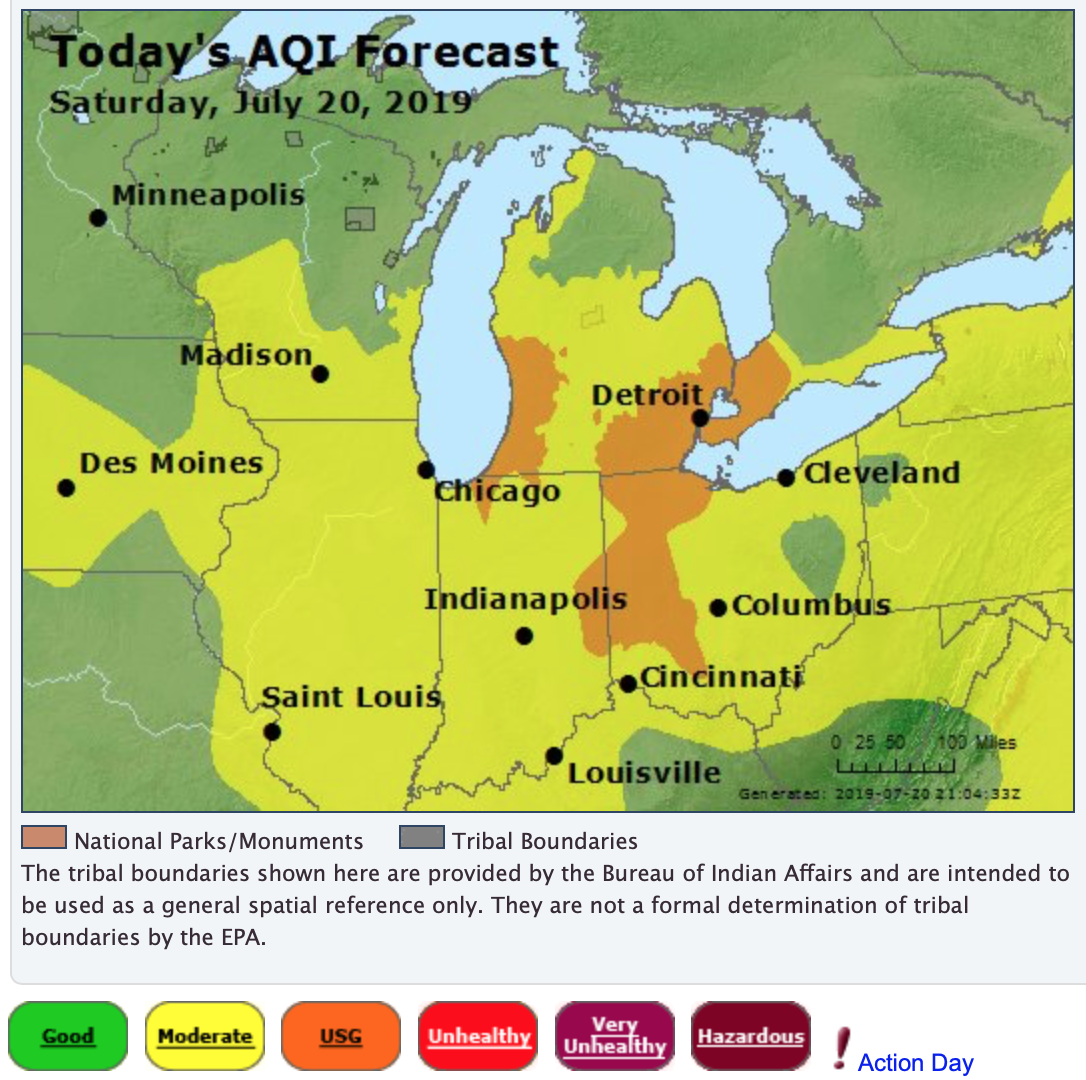}
        \caption{AirNow}
        \label{fig:airnow}
    \end{subfigure}
    \begin{subfigure}{0.5\columnwidth}
        \includegraphics[width=\columnwidth]{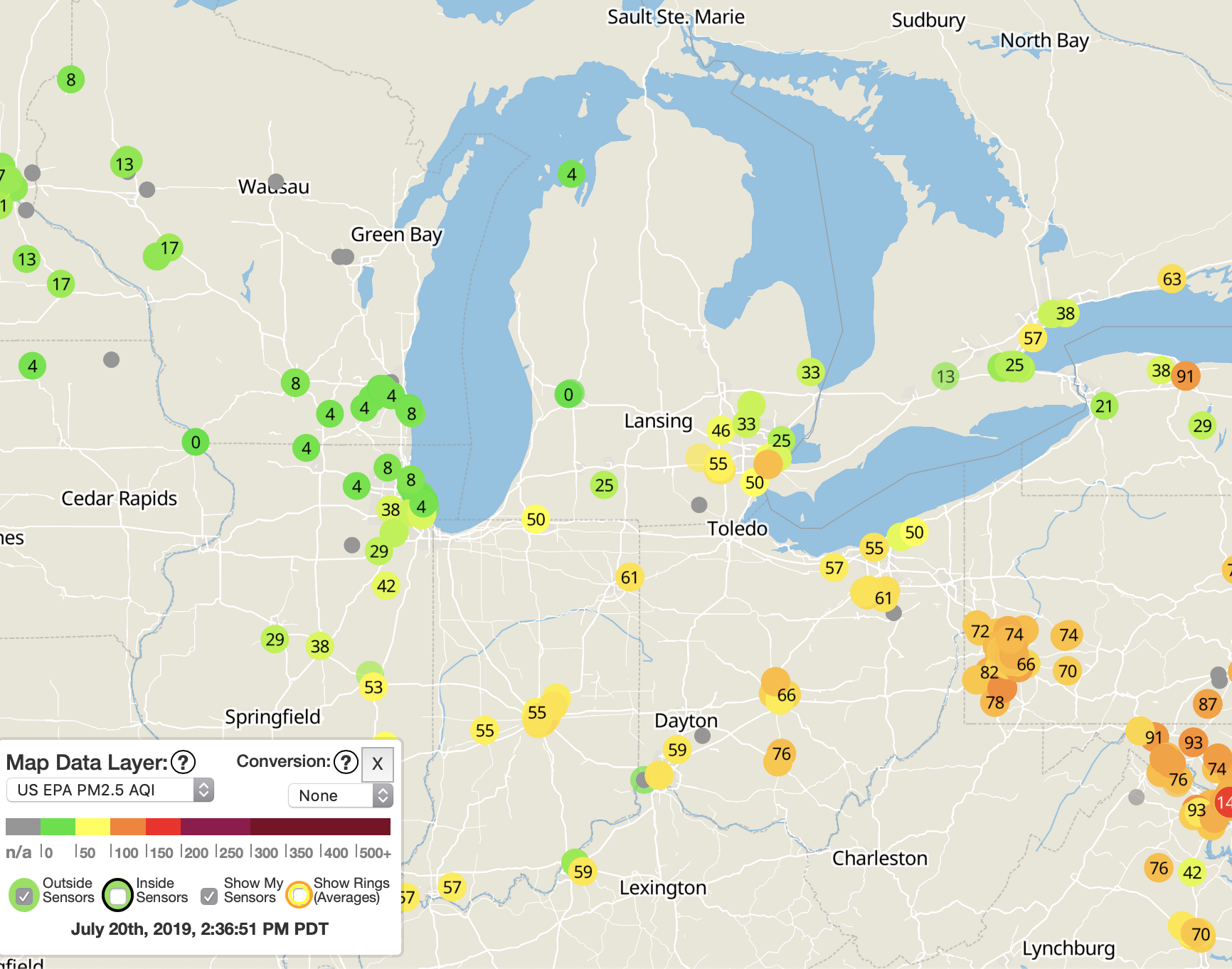}
        \caption{PurpleAir}
        \label{fig:purpleair}
    \end{subfigure}
    \caption{Common websites for checking air quality in the U.S.}
\end{figure}

Publicly available sources of air quality data exist worldwide. Many governments operate air quality sensors and, to varying degrees, make their data available online.
In the United States, the Environmental Protection Agency operates AirNow, a site showing a contour map colored by estimated air quality category (Figure~\ref{fig:airnow}). This estimate is an interpolation of the data from air quality sensors across the country.  
Colors indicate categories of health risk, each with corresponding guidelines (see Fig.~\ref{fig:aqi_legend}).

Recently, low-cost, internet-connected air quality sensors have become popular, such as those from PurpleAir. These sensors are installed by individuals in and around homes and buildings; their data are available online in real time. Low-cost sensors, offering better availability and wider spatial coverage of air quality data, could have a transformative effect on the public's awareness~\cite{perform_assessment}. Websites like IQAir and PurpleAir show estimates of air quality from these sensors. These visualizations may show contours, indicate sensor values directly as in Figure~\ref{fig:purpleair}, and/or show glyphs with aggregated sensor data. The interface gives a number summarizing the current pollution level in that location.

\subsection{Uncertainty}
In general, publicly available sources for air quality information do not include uncertainty. By neglecting uncertainty, predictions shown on sites like AirNow may often be inaccurate or misleading, and may underestimate air pollution in general. The raw data from sensors and the interpolation used to create a map are both sources of uncertainty, but users typically see no indication of either.

\begin{figure}[b]
\centering
\includegraphics[width=0.8\columnwidth]{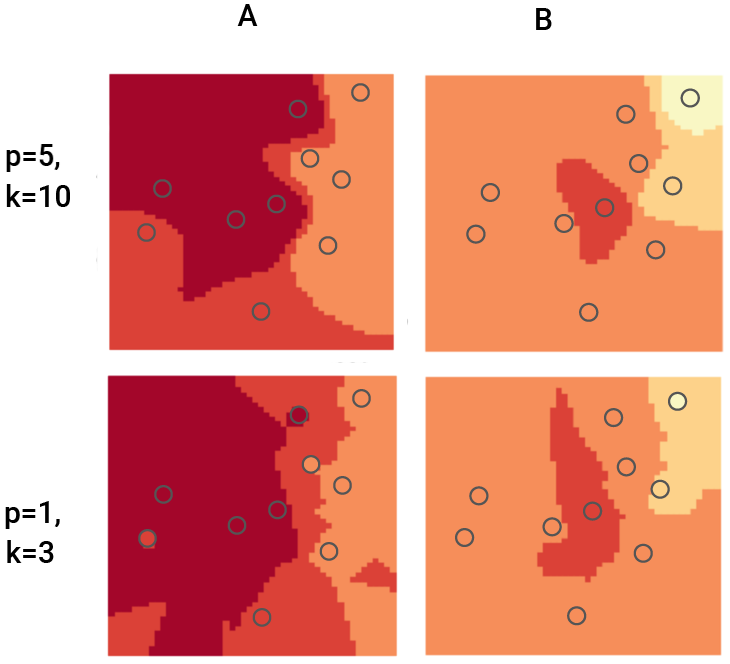}
\caption{Varying interpolation based on inverse distance weighting parameters for two sets of air quality sensor data (A and B). The parameters \textit{p=5, k=10} are used on AirNow, while \textit{p=1, k=3} are more commonly used for air pollution.}
\label{fig:idw}
\end{figure}

The placement of sensors is a primary source of uncertainty. For example, sensors are not evenly spaced across the U.S., and their distribution does not reflect the population distribution or the most significant pollution sources. For example, the interpolations shown on AirNow often do not capture variations on the scale at which pollution from traffic on highways is present, and sensors are often placed away from the population centers of cities. While government-owned sensors are helpful for monitoring long-term, large-scale air quality trends, they may not be offering completely truthful information in places where pollution is a chronic risk to public health. Air pollution from wildfires is also increasing drastically; many heavily affected areas are sparsely measured by the government's network of air quality sensors. Air quality estimates that users see, then, often do not reflect the most relevant sources of pollution and the potential variability in air quality.

Another source of uncertainty in air quality maps is the algorithm used to convert air sensor readings into a contour map. The contours shown on AirNow are based on Inverse Distance Weighting (IDW), a deterministic algorithm that predicts the value at a location by weighting detections from nearby sensors. IDW is widely used,
but its output is highly variable depending on the specific parameters used (see Figure~\ref{fig:idw}). The parameters used for the AirNow site differ significantly from parameters found to be optimal in other studies of air pollution interpolation~\cite{fontes_2010}. Techniques like cross-validation can be used to optimize the parameter values, but this is less effective when interpolating over large spaces, such as the vast distances between sensors in some parts of the United States. More sophisticated approaches exist (kriging is the most prominent alternative) but any approach requires assumptions and is prone to error. 

\begin{figure}[b]
\centering
\includegraphics[width=0.8\columnwidth]{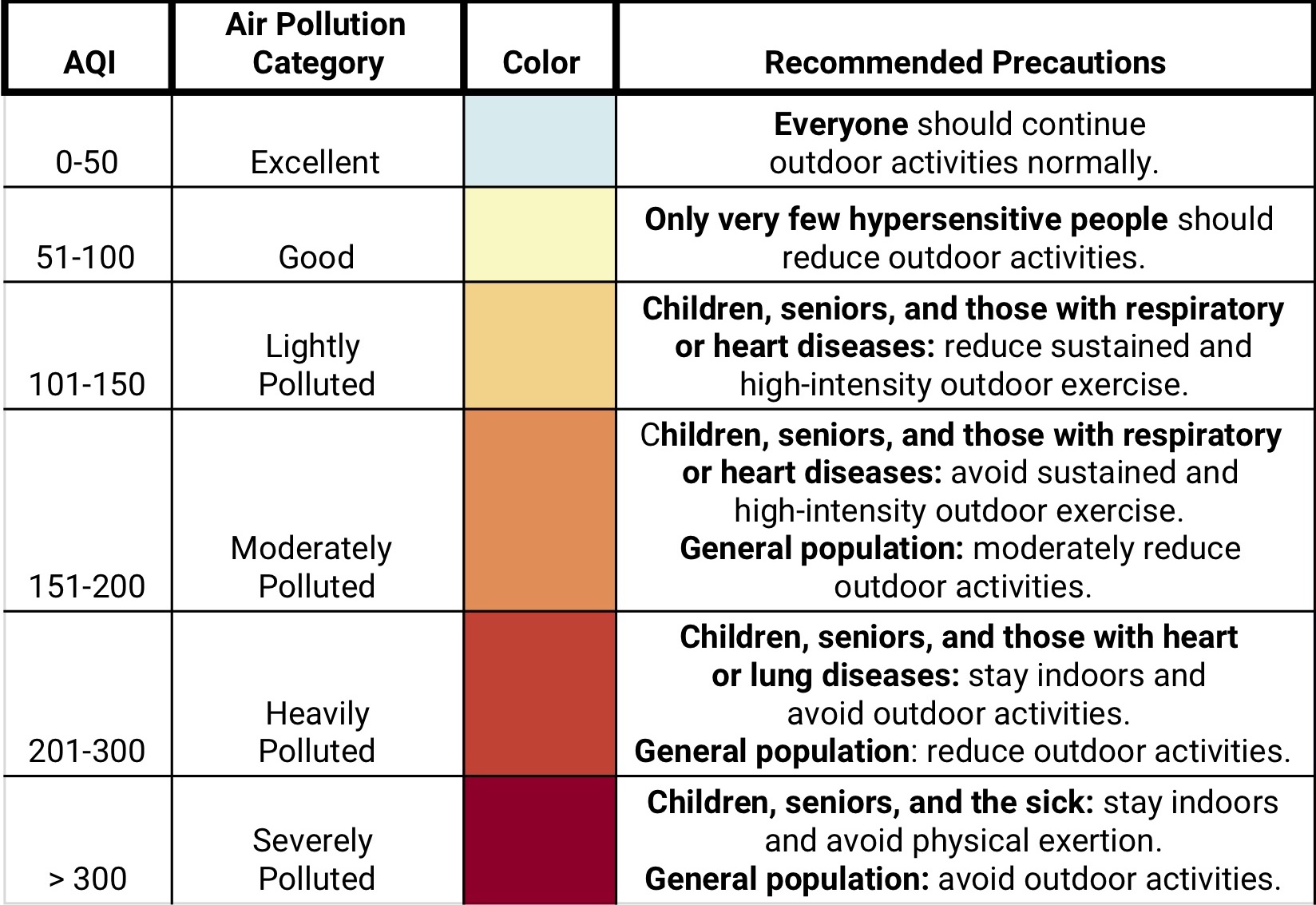}
\caption{Guidelines for each category of air pollution, from AirNow. Concentrations of a given pollutant are mapped to the Air Quality Index (AQI); AQI ranges correspond with certain health guidelines. For this study, we replace the default color map with a perceptually uniform map.}
\label{fig:aqi_legend}
\end{figure}

Though contour maps created from interpolations are widely used in geospatial communication, their uncertainty properties are not well-studied~\cite{contour-uncertainty}. Uncertainty in how sensors are labeled, or artifacts from their binning, have a significant impact on the resulting contours.
Researchers have proposed a method for identifying areas in choropleth maps where labeling has a strong effect on visual boundaries~\cite{choropleth_2017}. With this information, a map designer could adjust labeling to reduce potential bias. In the air quality case, guidelines exist for ranges of Air Quality Index (AQI) values (see Figure~\ref{fig:aqi_legend}), so a standard binning is already established. Maps need to be updated with evolving air quality readings, so manually adjusting maps is not feasible.

\subsection{Uncertainty Awareness and Decision-Making}
Research suggests that uncertainty information may help people make better decisions in their daily lives. For example, mobile transit apps showing uncertainty in bus arrival times may help people optimize when to leave for the bus~\cite{2016-when-ish-is-my-bus}. Including numeric uncertainty in weather forecasts may increase trust and help people make more holistic decisions~\cite{doi:10.1177/0963721413481473}, and the general public understands and expects uncertainty information in weather forecasts~\cite{Joslyn2010CommunicatingFU}. 

For risk communication in particular, research suggests that well-designed visual aids can be transparent, effective tools for un-biasing people's perception of danger~\cite{doi:10.1177/0963721413491570}. Some recent human-computer interaction work has focused on informing the public about underreported risks they face from flooding ~\cite{Soden:2017:TGL:3025453.3025983}. 
Uncertainty visualization enhances risk communication: users may be more willing to take appropriate precautions when shown an uncertain weather forecast compared to a categorical warning~\cite{Joslyn2012UncertaintyFI}. If an uncertainty-aware map shows the competing claims from air quality data sources,  people may be able to make more informed decisions and understand the sources and characteristics of air pollution in their city compared to others.


\subsection{Survey: Air Quality Awareness}
\label{aqa}
 One challenge in characterizing people's understanding of air quality information is that many people in the United States may not regularly engage with it. One study found that about 12\% of the population had changed their behavior within the past year in response to poor air quality, but those with respiratory conditions are more likely to take action~\cite{wells2012}. We conducted an online survey to gauge people's awareness of local air quality, their sources of information, and their responses to poor air quality.

We decided to focus on users who already have some awareness of air quality, in order to yield respondents who are motivated to pay attention to their health and may have taken preventative action in the past. We targeted people 18 and older in our local community, which had experienced significant air pollution from wildfires within the past year. We had 54 respondents, 40 of whom fully completed the survey. Of the respondents who indicated being aware of poor air quality within the past year ($n=34$), $91\%$ specified that this unhealthy air was due to a wildfire.

In general, our survey indicated that people aware of air quality issues are interested in simple AQI information and are likely to make changes and trust sources of information, but may not have a consistent rationale or source for these decisions. When asked whether they feel they have accurate, complete information about the air quality near them, $72\%$ ($n=39$) at least somewhat agreed. 

We asked respondents whether they had used any websites in the past 12 months to check air quality; $78\%$ indicated they had. Of those who did, $35\%$ used AirNow alone, $32\%$ used another site alone, and $32\%$ used a combination of AirNow and another site. At least $67\%$ of respondents, then, use a contour-based air quality map.

Most respondents ($91\%$, $n=34$) indicated that they had changed their behavior within the past year due to poor air quality. The changes these respondents have made are often significant: $90.3\%$ of those people said they had avoided time outside, $74\%$ cancelled or skipped activities, and $65\%$ did less strenuous exercise.



\section{Methodology}
Our goal is to test visualization designs that might help improve people's decision-making, align with their desires and needs for air quality tracking, and accurately represent risk. To assess the potential of these designs in realistic settings, we use real air quality data representative of what is currently available. We also aim to quantify the uncertainty in the data in a way that is representative of actual variation.

\subsection{Sources of Data} 
In this study, we focus on the harmful pollutant PM2.5, which comes from automobiles and wildfires, among other sources. 
AirNow is considered the ground truth in the U.S. for air pollution readings, but the sensor coverage is limited in many areas of the United States, and in more densely populated states, sensor coverage is disparate from the distribution of people and sources of pollution (e.g. roads).




PurpleAir is one of the most prominent sources for low-cost air quality sensors. Individuals around the world can purchase sensors to install inside or outside homes or other buildings (we consider only outdoors sensors here). These sensors are connected to the internet and their data are made available at purpleair.com in real time. 
Because PurpleAir sensors are not maintained by the company once they are installed by individuals, their accuracy may worsen over time. In particular, dust and other debris accumulates over the laser-based sensors, and without cleaning, the readings may drift. These sensors do not directly measure PM2.5 concentration, instead inferring it from other measurements.





\subsection{Quantifying Uncertainty}


One source of uncertainty is the interpolation approach, as described in section 1.1.1. Two common approaches in geospatial applications are inverse distance weighting (IDW), which is deterministic, and kriging, which is probabilistic. IDW, used for AirNow, predicts values at points by taking a weighted average of the $k$ nearest neighbors.  The results of IDW are highly dependent on the parameters used, $k$ and $p$, where the $k$ nearest sensors are weighted by their distance to a location, $\frac{1}{(\mathrm{dist}^p)}$.

Kriging is a statistical approach based on characterizing the autocorrelation between pairs of detections. This approach requires more tuning than IDW, such as specifying the shape of the semivariogram describing the autocorrelation. Previous work examines the use of kriging vs. IDW for interpolating air quality, finding that which method is superior depends highly on the scenario and on the particular pollutant~\cite{fontes_2010}. In this study, we use kriging to generate visualizations because of its more natural relationship with uncertainty; kriging algorithms estimate a mean value and standard deviation at each grid point.

\section{Visualization Design}
We explored possible designs of static uncertainty-aware air quality visualizations to compare with standard contour- or sensor-based designs. For perceptual uniformity across the AQI scale, we use an updated color map (Figure~\ref{fig:aqi_legend}) reminiscent of the AirNow version (Figure~\ref{fig:airnow}). This map might contribute helpful associations: one study suggests that people choose darker colors in ``negative'' and ``disturbing'' color schemes~\cite{affective_color}. In another study, people associating darker colors with “more” (more pollution in this case) may have had a stronger influence on risk belief than other factors such as number of colors and level of focus used, leading the authors to conclude that “incrementally darker shading was very effective for conveying incremental risk"~\cite{Severtson2013}.

We chose designs along a spectrum of uncertainty awareness. Standard designs (Section~\ref{non_uncertainty}) show no uncertainty (\textit{interpolation only}) or show it implicitly (\textit{interpolation with sensors}). Uncertainty-aware designs (Section~\ref{uncertainty_views}) involve explicitly encoding uncertainty with either 2 (\textit{risk contour map}) or 9 (\textit{small multiples}, \textit{dotmaps}) possible outcomes. Previous work~\cite{Greis:2018} studied how visualizations representing different amounts of uncertainty information influence the way users interpret data aggregated from sensors. Their research suggests that the amount of uncertainty shown affects our mental models for interpreting data. We expect, then, that our designs might elicit different types of reasoning. In these designs, we consider only the uncertainty due to the non-systematic locations of air quality readings.

\subsection{Standard Views}
\label{non_uncertainty}

\begin{figure}[h]
\centering
        \includegraphics[width=\columnwidth]{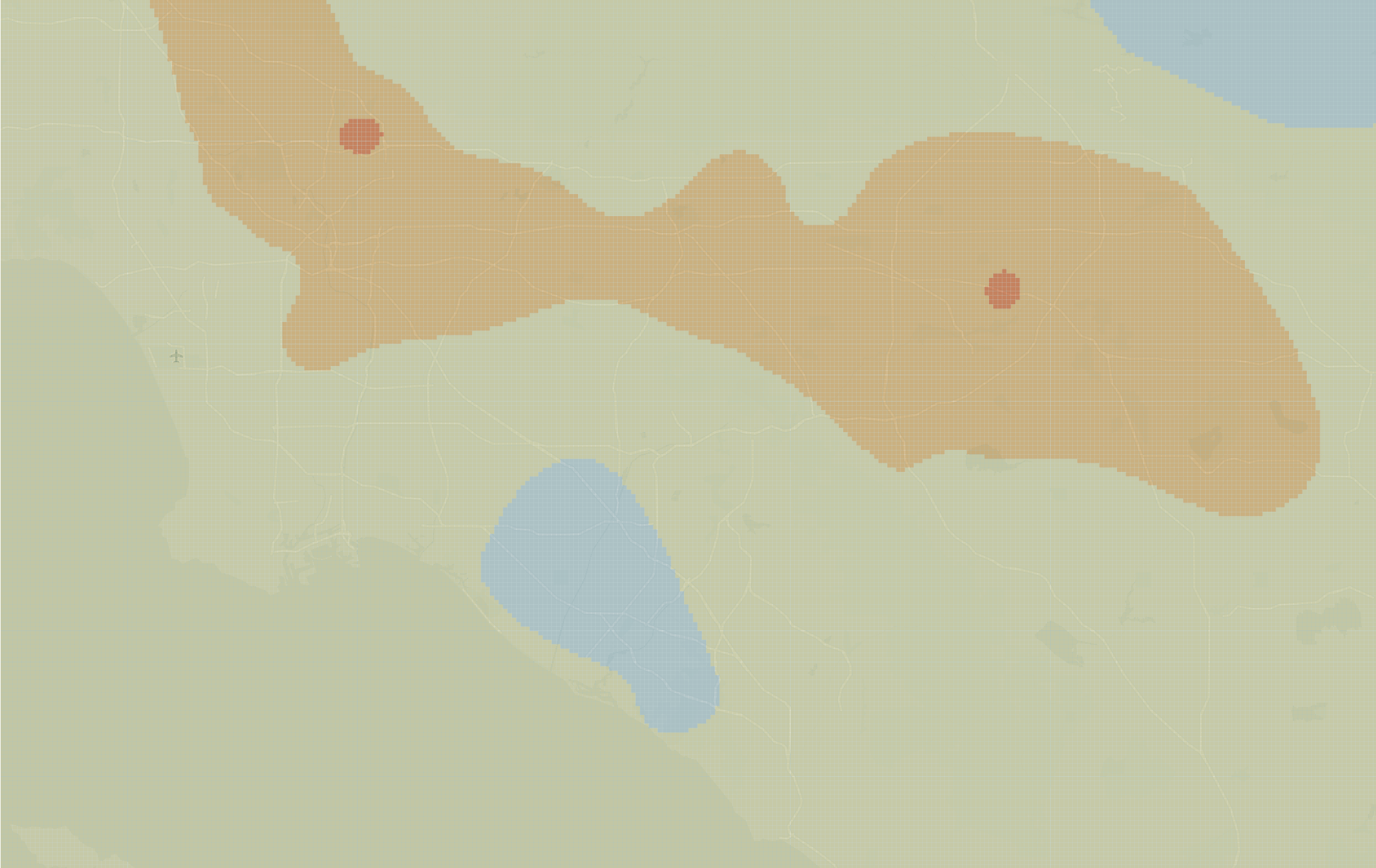}
        \caption{Interpolation only. The contours represent one estimate based on sensor data, with no uncertainty information.}
        \label{fig:interp_only}
\end{figure}

\textbf{Interpolation only} (Figure~\ref{fig:interp_only}) shows no uncertainty, representing the status quo in air quality visualization. Contours are based on the mean interpolated kriging estimate from sensor detections (see Section~\ref{uncertainty_views}). We include this view as a baseline to understand typical reasoning.

\begin{figure}[h]
\centering
        \includegraphics[width=\columnwidth]{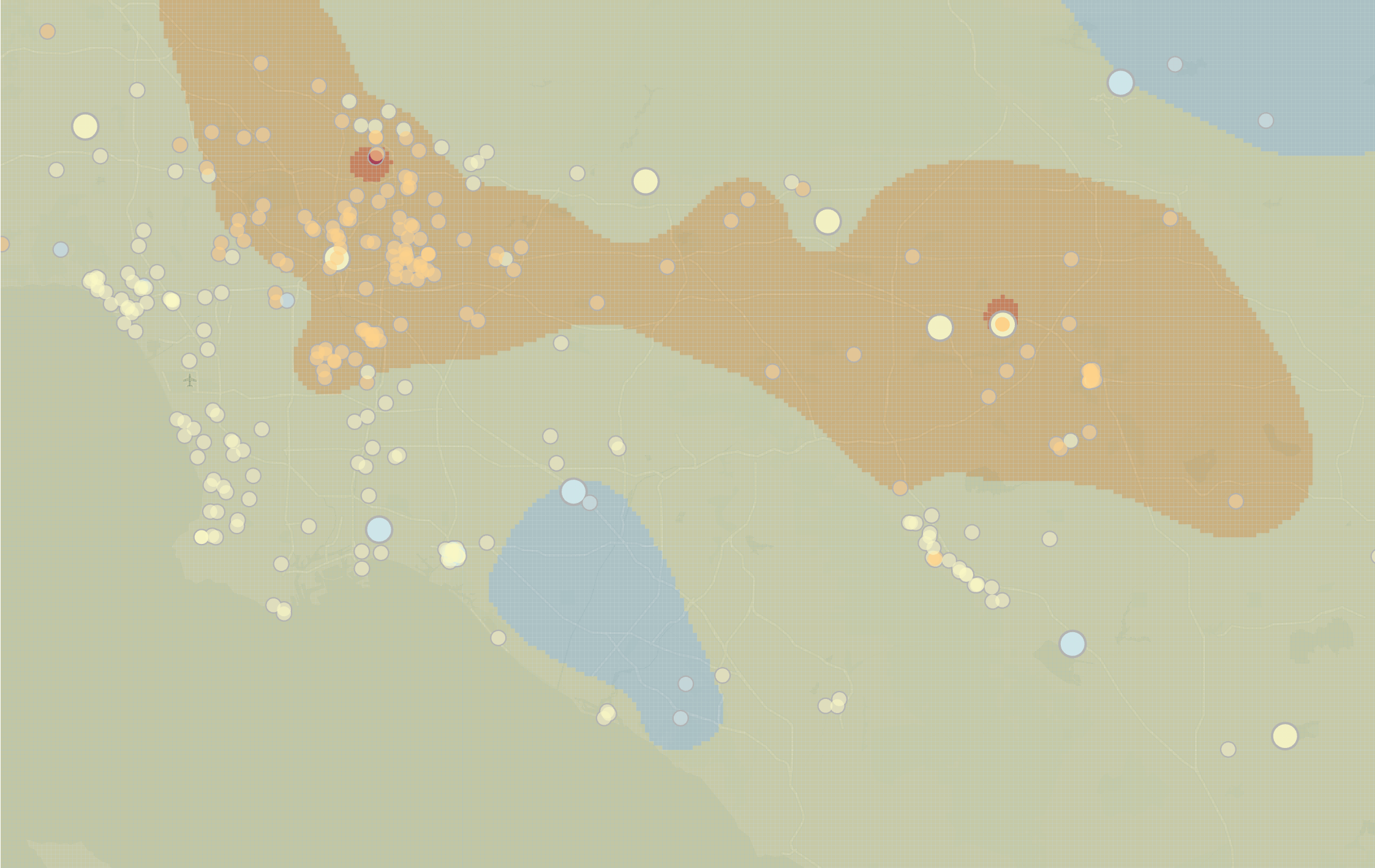}
        \caption{Interpolation with sensors. Uncertainty is implied by the conflict between sensor data and the best-fit interpolation (larger=government-owned; smaller= PurpleAir).}
        \label{interp_w_sensors}
\end{figure}

\textbf{Interpolation with sensors} (Figure~\ref{interp_w_sensors}) represents a standard type of visualization available on air quality monitoring sites. (Often, these views only show sensors and do not include interpolation; we left that case out of this study.) Research suggests that users prefer to be able to view conflicting individual data sources even when an aggregation---in this case, the contour map---is available~\cite{Greis:2017}. We indicate relative reliability by encoding the government-owned sensors with larger circles.

This view lets us ask how people's understanding of an interpolation changes if the underlying information is also shown.  Discrepancies between the raw data and the chosen interpolation are an implicit representation of uncertainty. Conflicts between different sensors show users some ambiguity, perhaps encouraging thought about how the interpolation was derived from the data, especially when the measurements seem disparate from the interpolation. 

Prior work suggests how users might interpret these maps: people may aggregate information differently if given access to uncertainty information, weighing each source of information and taking its reliability into account~\cite{Greis:2018}. Specifically, users are likely to mentally average the sensor information together to reach a conclusion, maybe using a weighted average if the sensors have different reliability.

\subsection{Uncertainty Views}
\label{uncertainty_views}
Researchers have proposed thinking of a set of possible outcomes in an uncertain situation as \textit{multivalued} data~\cite{1438260}. Distinct from \textit{multivariate} or \textit{multidimensional} data, in \textit{multivalued} data, each datum has a collection of values for a single variable (in our case, possible AQI values at each location). The authors pointed out that few  geospatial visualizations had treated uncertainty as multivalued data without using animation. Multiple linked displays have been used for exploring ensembles of outcomes of simulated geospatial data~\cite{potter:2009:EVSV}. In the interest of public accessibility and distribution potential, however, we limit ourselves to  static views. Integrating uncertainty into the map itself is likely to be more influential for people's decisions than providing an adjacent uncertainty view, and it may be easy to ignore uncertainty information presented separately~\cite{2017-trust-but-verify}.

Significant work in geospatial visualization has focused on techniques that can be integrated into static views, including textures, transparency, hue, and value~\cite{kinkeldey_2014_survey}. For example, bivariate color maps have been proposed to integrate data and uncertainty into one image~\cite{Lucchesi_Wikle}. A variant of this idea is the Value-Suppressing Uncertainty Palette (VSUP); encoding information and uncertainty in a VSUP encourages reasoned, uncertainty-aware decision making~\cite{2018-uncertainty-palettes}.

In general, these approaches are more suited to showing the magnitude of uncertainty rather than depicting the relative probabilities of possible outcomes. Texture and color-based approaches work by downplaying or obfuscating more uncertain information. Their underlying premise is that differences among more certain data are more important than differences among highly uncertain data~\cite{2018-uncertainty-palettes}. 

Another approach to uncertainty visualization is to fairly represent the relative likelihoods of different outcomes. Recent work in uncertainty visualization has shown promise in \textit{direct displays} of these ensembles of potential outcomes. For example, showing a sampled ensemble of hurricane predictions can improve users' ability to estimate danger over \textit{summary displays}, which depict the mean and its spread~\cite{rep_sampling},~\cite{hurricane_tracks}. In this study, we focus on transferring direct displays into a map context. We considered a wide range of designs that could show direct displays of uncertainty, ultimately using the following four designs.

\begin{figure}[t]
\centering
\includegraphics[width=\columnwidth]{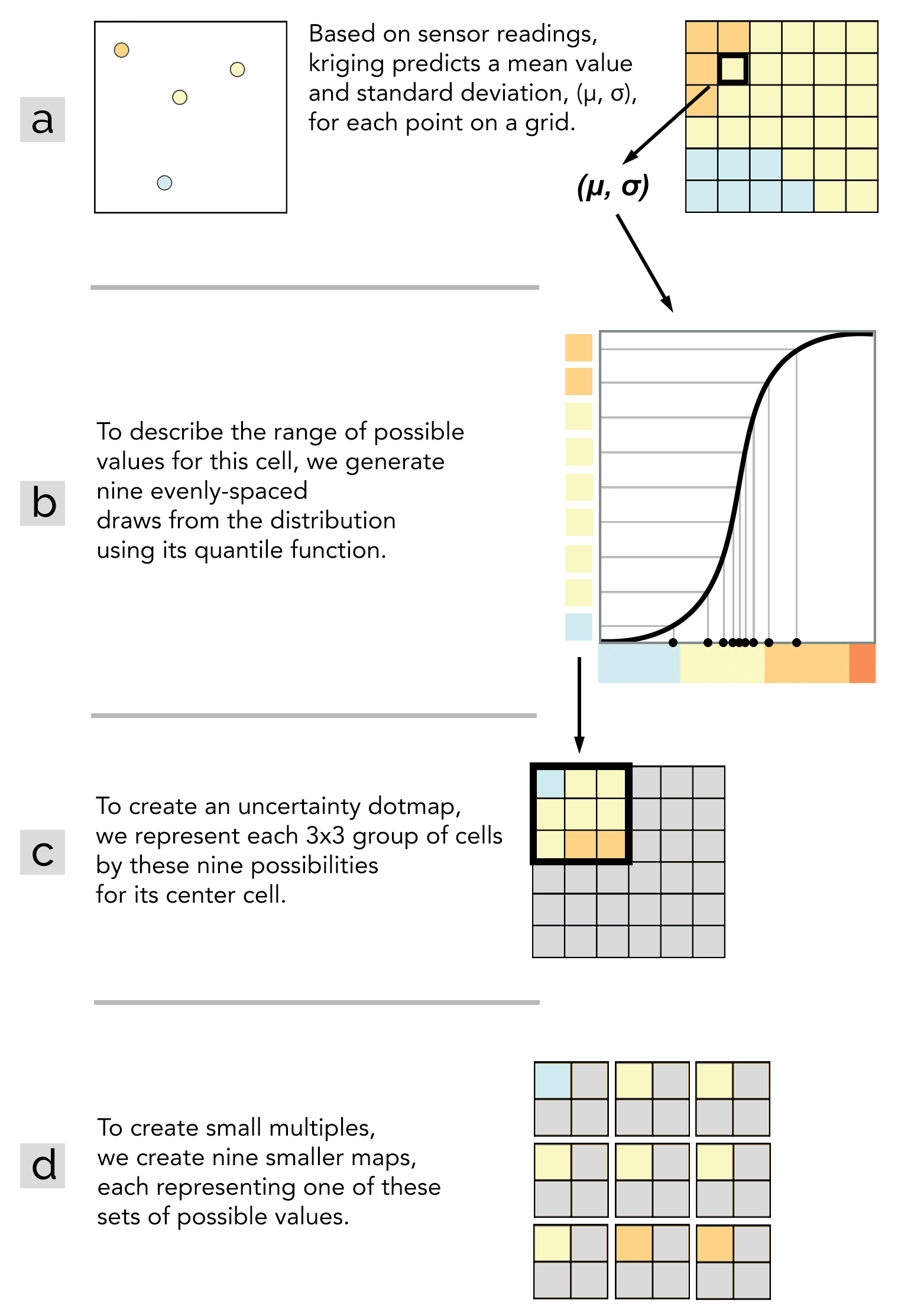}
\caption{Creating small multiples and dotmaps from kriging results. Uncertainty information is used to create an ensemble of nine outcomes for each set of air quality readings.}
\label{fig:dotmap_explanation}
\end{figure}
To quantify the uncertainty information underlying these views, we use kriging to describe a range of possible outcomes from measured sensor data. Figure~\ref{fig:dotmap_explanation} shows how we sample this kriging grid to convey uncertainty for different visual designs. The mean kriging-based estimate for each grid cell is used to create the contours in the standard views in Section~\ref{non_uncertainty}. 
For the \textit{interpolation only} option---showing a contour map without any uncertainty---the interpolated visualization has a higher resolution, i.e., it samples more points on the kriging grid. When we add in uncertainty information, we sacrifice some of the space available for showing the mean estimate in exchange for more information about the standard deviation.

\begin{figure}[h!]
\centering
      \includegraphics[width=\columnwidth]{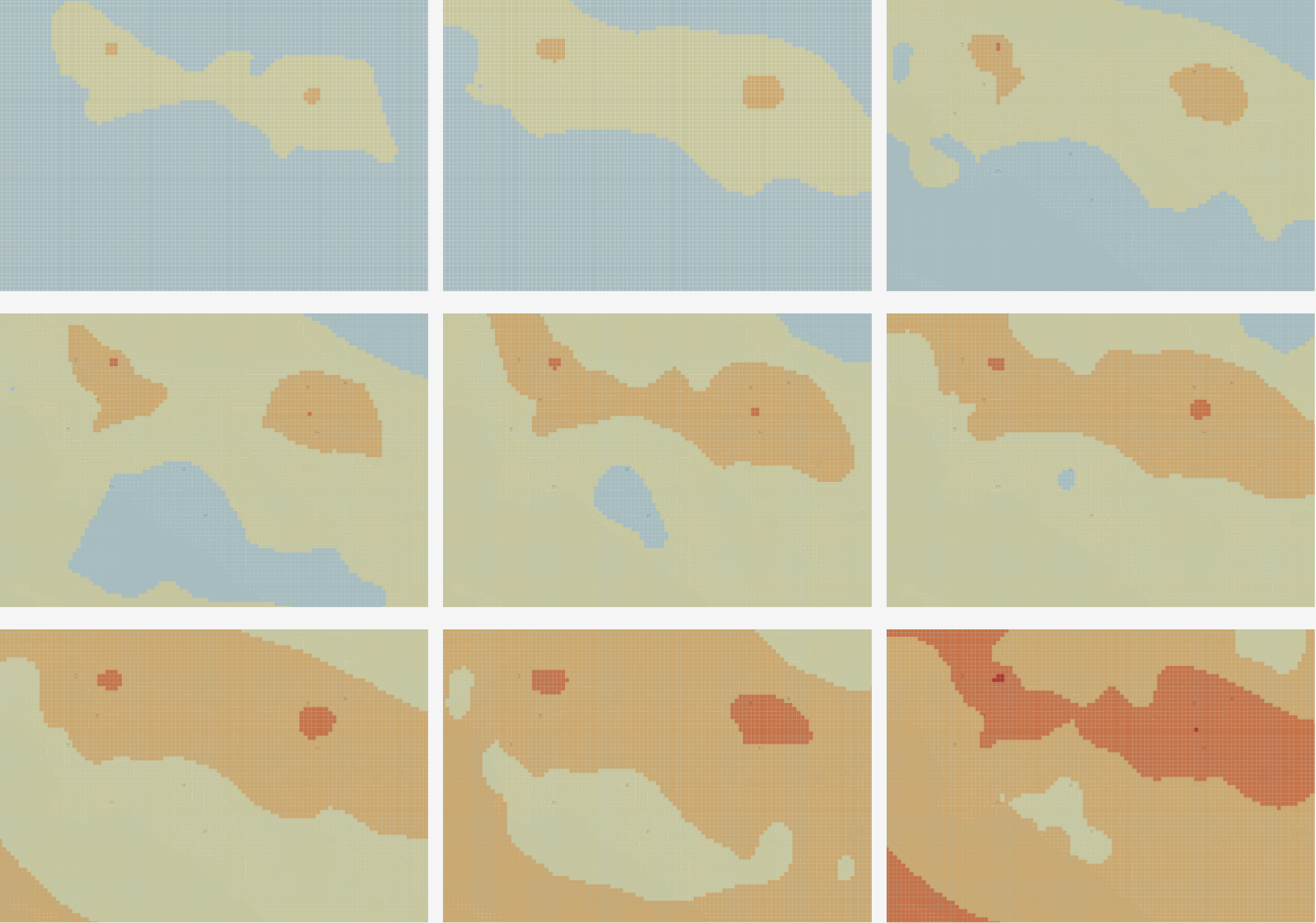}
      \caption{Small multiples. Uncertainty is depicted using nine different possible interpolation outcomes given the same air quality data, ordered from best-case to worst-case scenario.}
      \label{fig:small_multiples}
\end{figure}  
\textbf{Small multiples} (Figure~\ref{fig:small_multiples}) align with current research in uncertainty visualization. They are a frequency-framing way to understand the uncertainty inherent in an estimate, showing the different possibilities all at once. Showing uncertainty with discretized outcomes may improve user recall~\cite{2018-imagining-replications} and help with confident, optimal decisions~\cite{2016-when-ish-is-my-bus}.

In general, small multiples are not used to convey discretized uncertainty, though encoding comparisons of ``multiple realizations'' in geospatial uncertainty visualization has been proposed~\cite{maceachren:2005:VGIU}. In our case, it is a way to show uncertainty via a direct display of possible outcomes. We propose that this view might help users make optimization judgments like whether to reduce their outdoor activity.

Previous work in visualizing multiple kriging results of air quality data suggests that interactivity is vital, for users to see how the probabilities change according to threshold~\cite{pebesma:2007:IVUS}. Without interactivity at our disposal, small multiples capture representative snapshots reflecting this type of reasoning, showing the map at each of nine thresholds. Note that in our case, the small multiples can be ordered from most optimistic to most pessimistic scenario, while it is not always possible to order uncertain outcomes this way.

\begin{figure}[h]
\centering
        \includegraphics[width=\columnwidth]{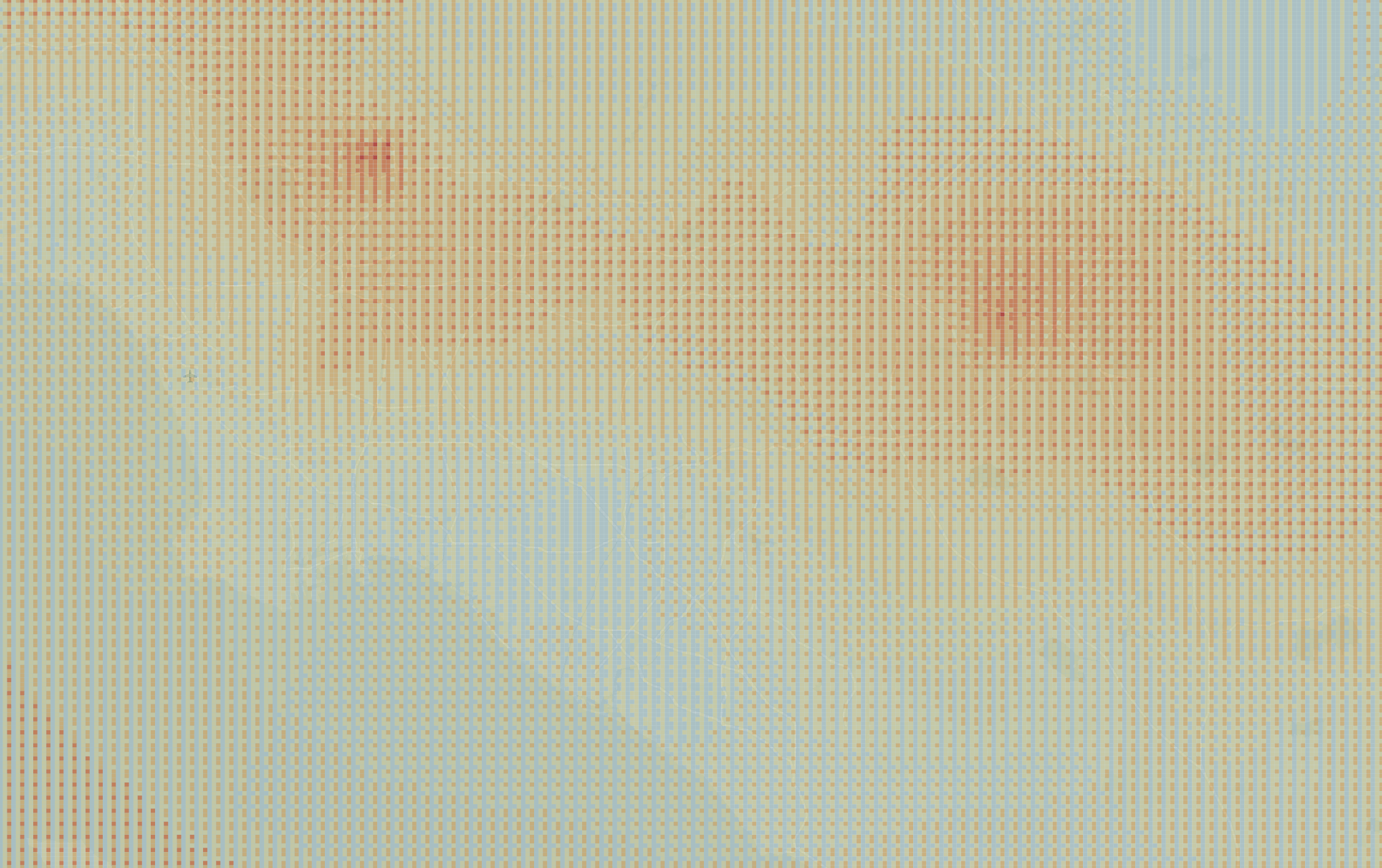}
        \caption{Ordered dotmap. The nine possibilities depicted in the small multiples are overlaid here onto a single map, with each 3x3 group of cells depicting nine possible outcomes.}
        \label{ordered_dotmap}
\end{figure}

\begin{figure}[h]
\centering
        \includegraphics[width=\columnwidth]{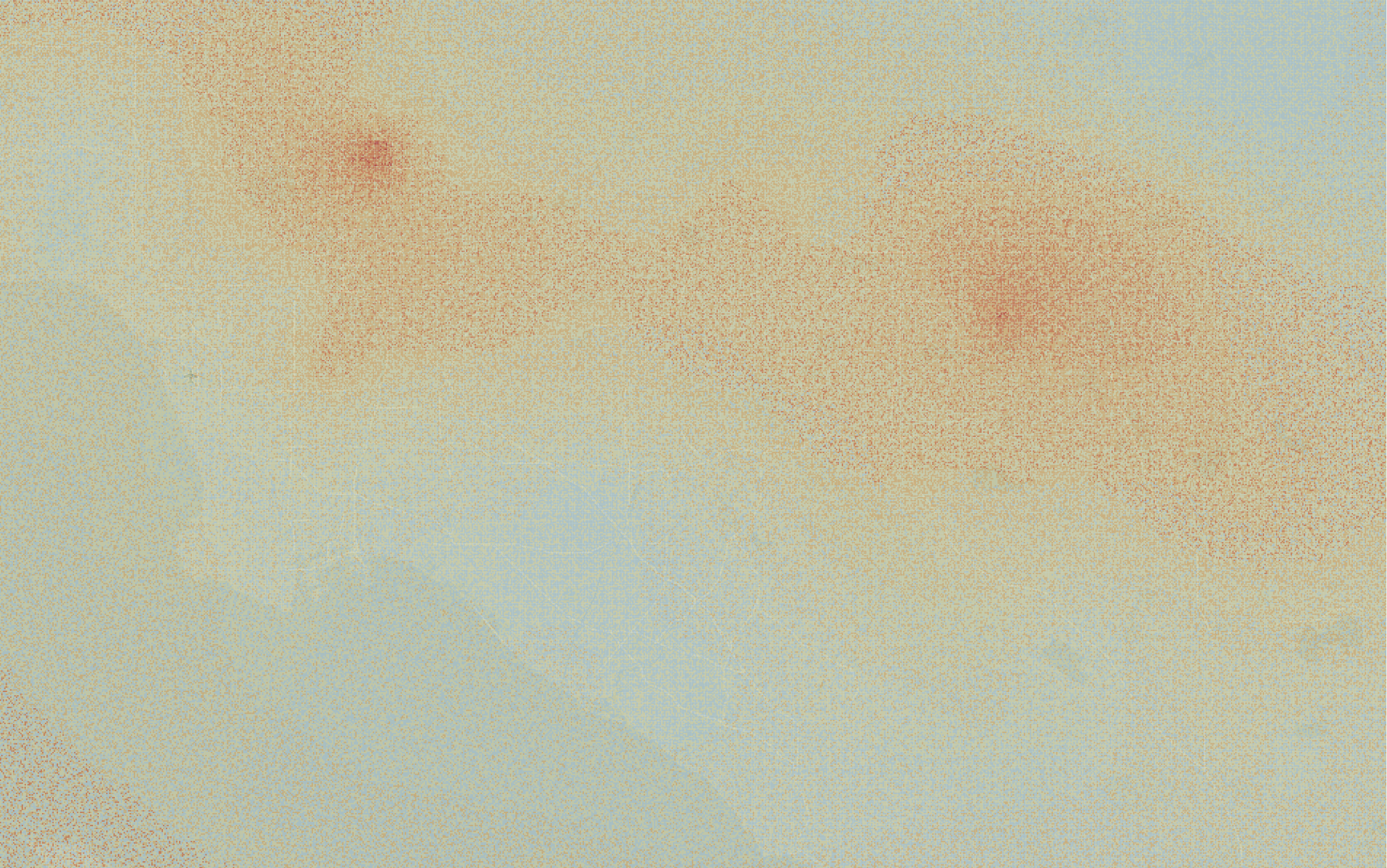}
        \caption{Smoothed dotmap. The information is the same as in the ordered dotmap, but each cell is smaller and the order of the colors is randomized within each set of cells.}
        \label{smoothed_dotmap}
\end{figure}

\textbf{Dotmaps} (Figures~\ref{ordered_dotmap},~\ref{smoothed_dotmap}) are a way to show a discretized representation of uncertainty at each point on the map. We use groups of colored grid cells to represent the distribution of possible air quality estimates at that location (see Figure~\ref{fig:dotmap_explanation}). This idea was inspired by dotplots~\cite{2016-when-ish-is-my-bus}. 

Similar static techniques have been proposed, using pixelation on maps to convey uncertainty. Building on previous ideas including using texture and flickering pixels to convey uncertainty, researchers have proposed a pixelated choropleth map to convey uncertainty of a value within counties~\cite{Lucchesi_Wikle}. Using a monochromatic scale, pixels are assigned colors based on random draws within the margin of error. 


We include two different dotmaps in our study. One is an ``ordered'' dotmap, with cells large enough to discern individually, and ordered within their groups. This might encourage users to compare relative frequencies of colors in different areas of the map. The second version is the ``smoothed'' dotmap. To create these, we transform each 3x3 group of cells into a 9x9 group of smaller cells with the same percentage of cells per color. These smaller cells are each placed randomly within the outline of the original 3x3 grid. Smaller grid cells may encourage users to visually interpolate the colors to come up with intermediate values. It may be difficult to discern values in highly uncertain areas, since they will look noisy. Previously proposed uses of texture and pixelation for uncertainty in maps sometimes have this goal of obfuscating more uncertain information~\cite{Lucchesi_Wikle}, or using lack of focus to suggest uncertainty~\cite{Severtson2013}.

In addition to encoding the ensemble of estimates, dotmaps may also reduce the appearance of boundaries, which have a strong effect on people's perception of uncertain map data~\cite{choropleth_2017}. Reducing firm borders may encourage users to think more about uncertainty~\cite{Severtson2013}. We also want to see if people can use the ordered dotmaps to interpret probabilities, or relative frequencies of outcomes. A similar approach such as stippling may allow visualization design that more finely tunes the tradeoff between clear borders and local details~\cite{Goertler2019StipplingScalarFields}.

\begin{figure}[h!]
\centering
        \includegraphics[width=0.94\columnwidth]{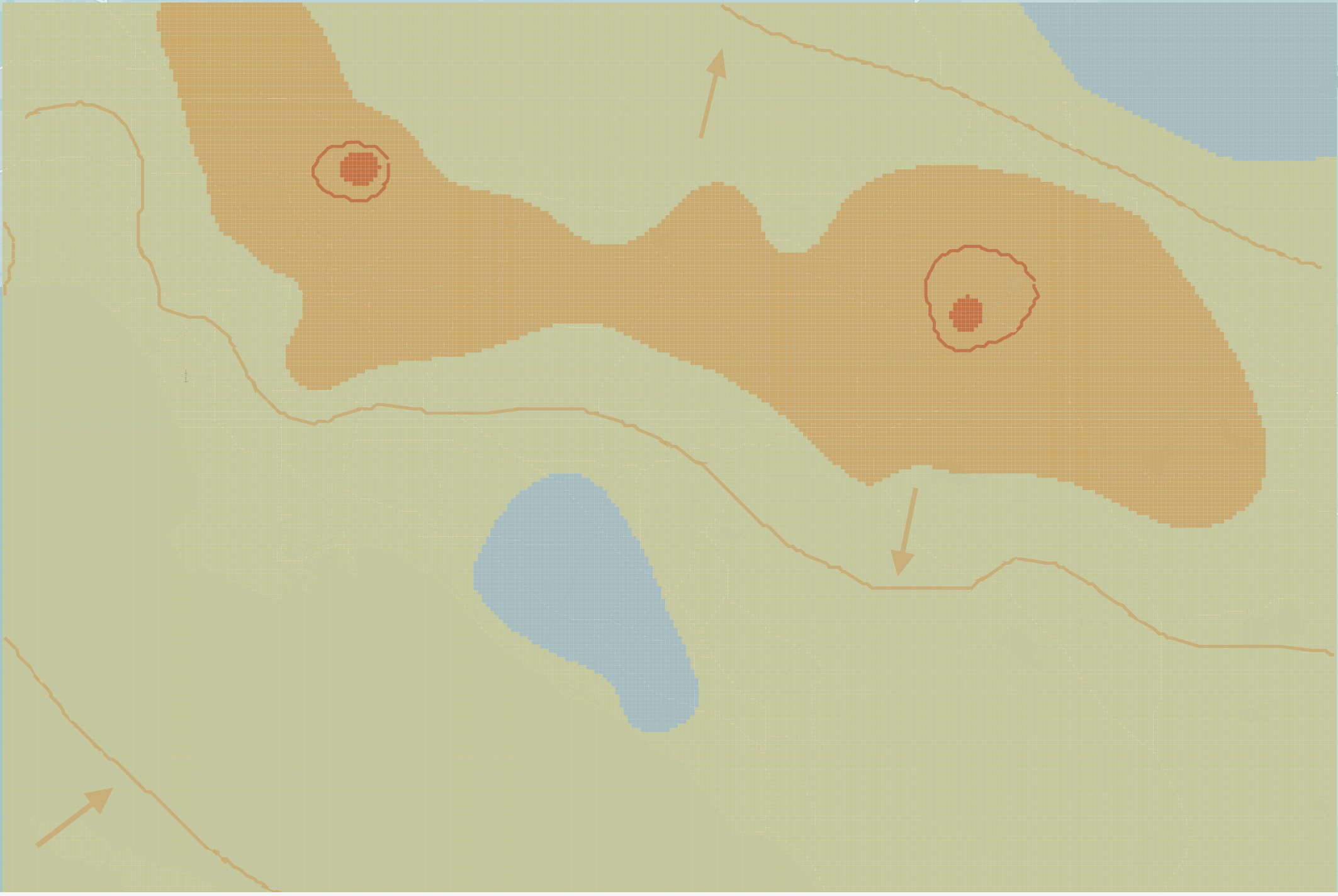}
        \caption{Contour map. The most likely interpolation is shown, overlaid with contours and arrows depicting one worst-case estimate.}
        \label{risk_contours}
\end{figure}

\textbf{Risk contour maps} (Figure~\ref{risk_contours}) are a hybrid approach, emphasizing the default estimate but highlighting the possibility of the $75^{\textrm{th}}$ percentile. This option presents a less ``fuzzy''-seeming view of uncertainty. The discrete boundaries in contour-based maps may have a strong impact on how  air quality is perceived; users may be judging the significance of different air quality regions based on the size of each area~\cite{klippel_2011}. 

To create the maps, we show the median estimate map, and overlay isocontours from the map of the $75^{\textrm{th}}$ percentile estimate. Due to some ambiguity in the contour shapes, we include arrows in these visualizations to indicate the direction of worsening prediction. (For example, in Figure~\ref{risk_contours}, the median estimate for the orange area is shown, while the $75^{\textrm{th}}$ percentile estimate for this area outlined in orange; there is a chance the orange area might be as large as the outlines.) This view shows less uncertainty information than the small multiples or dotmaps, depicting two possible estimates rather than nine. Contours are a familiar representation, so using them to encode areas of heightened risk may be intuitive and help acclimate users to thinking about uncertainty. However, depicting uncertainty by adding discrete boundaries to a map may be misleading by drawing attention to the particular border placement~\cite{choropleth_2017}.





\begin{figure*}[h!]
\centering
    \begin{subfigure}{0.45\textwidth}
        \centering
        \includegraphics[width=\textwidth]{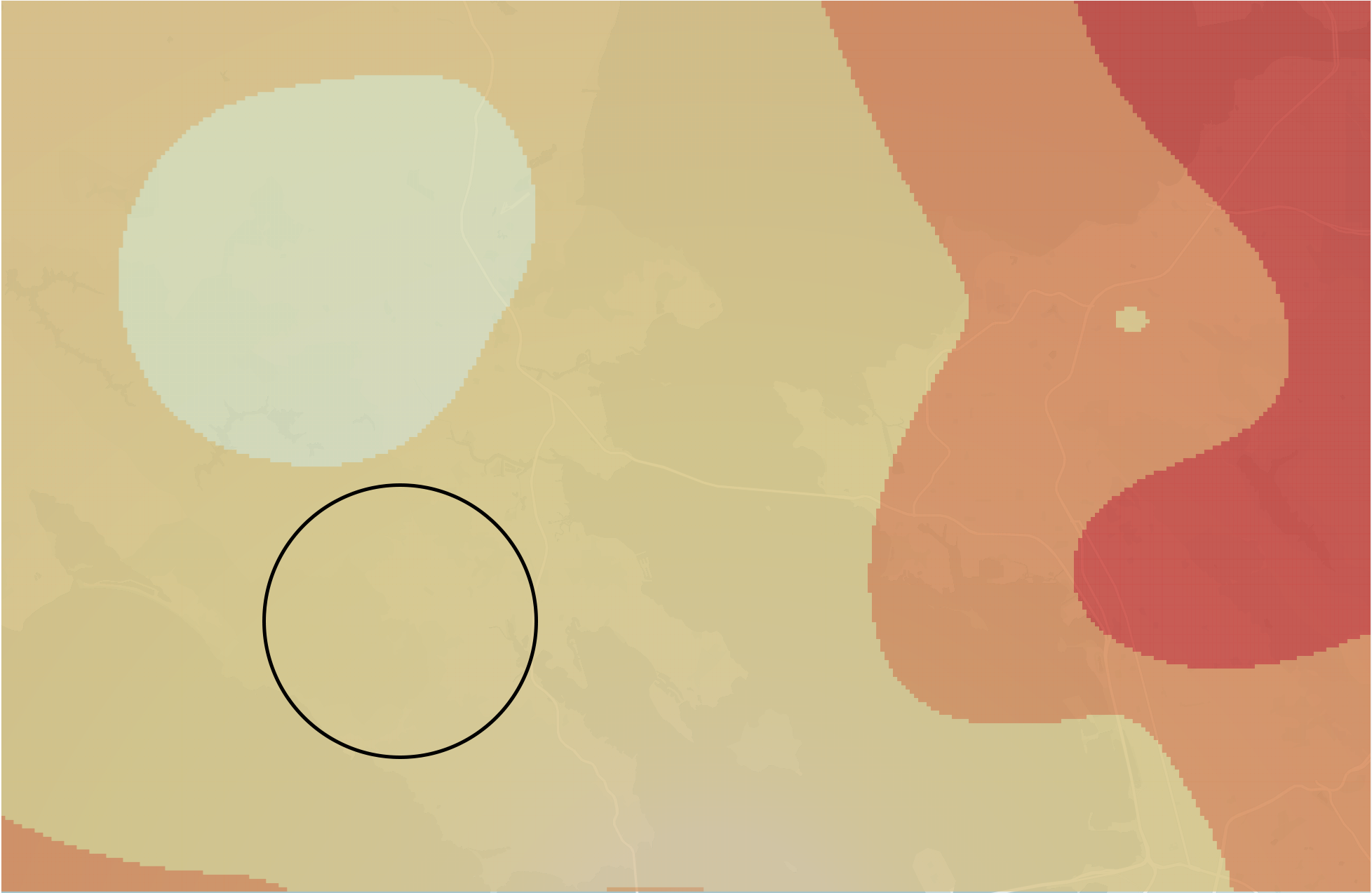}
        \caption{Interpolation only.}
        \label{interp_only}
    \end{subfigure}\hfill%
    \begin{subfigure}{0.45\textwidth}
        \centering
        \includegraphics[width=\textwidth]{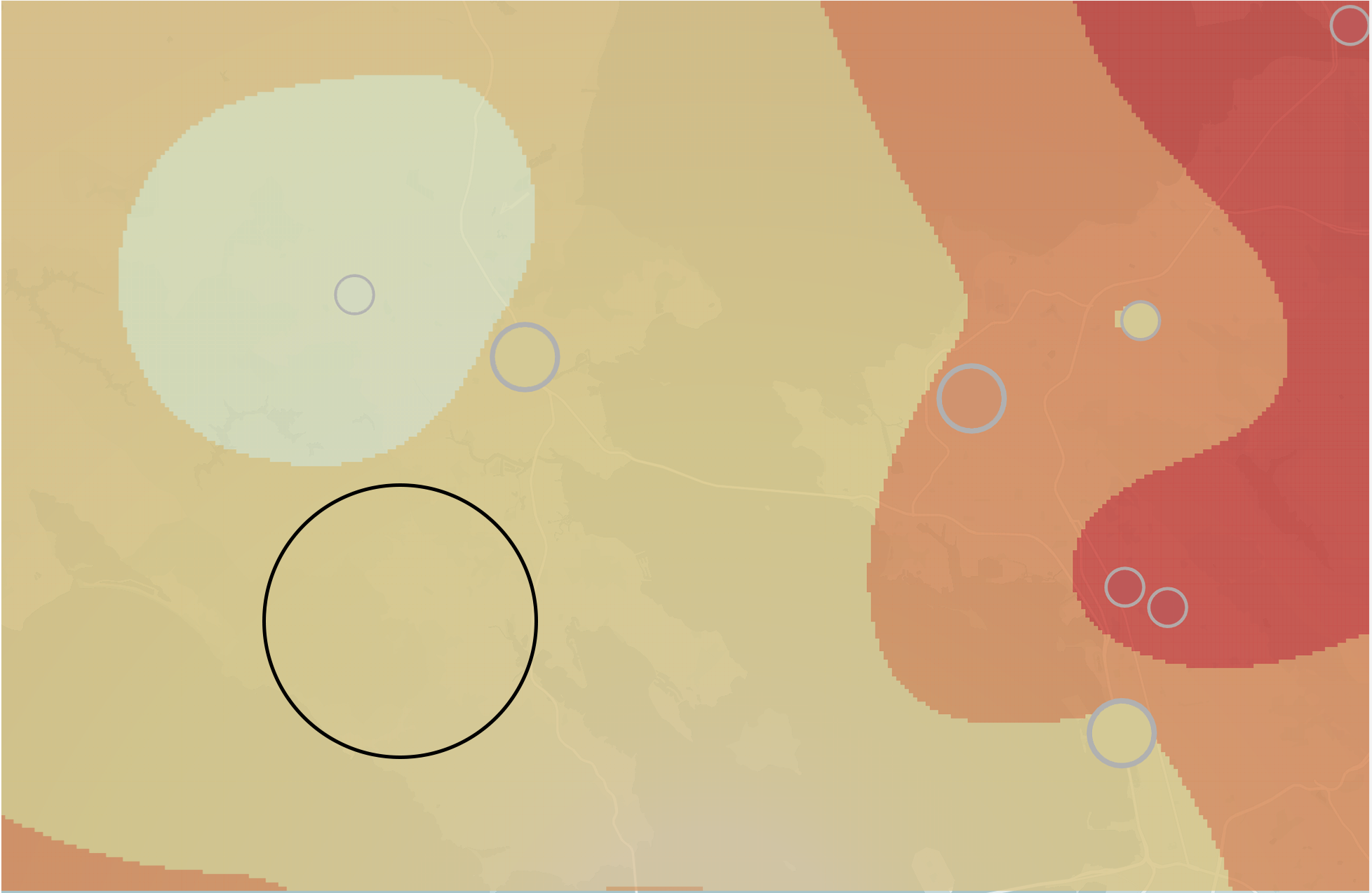}
        \caption{Interpolation with sensors.}
        \label{interp_and_sensors}
    \end{subfigure}\hfill%
    \begin{subfigure}{0.45\textwidth}
        \centering
        \includegraphics[width=\textwidth]{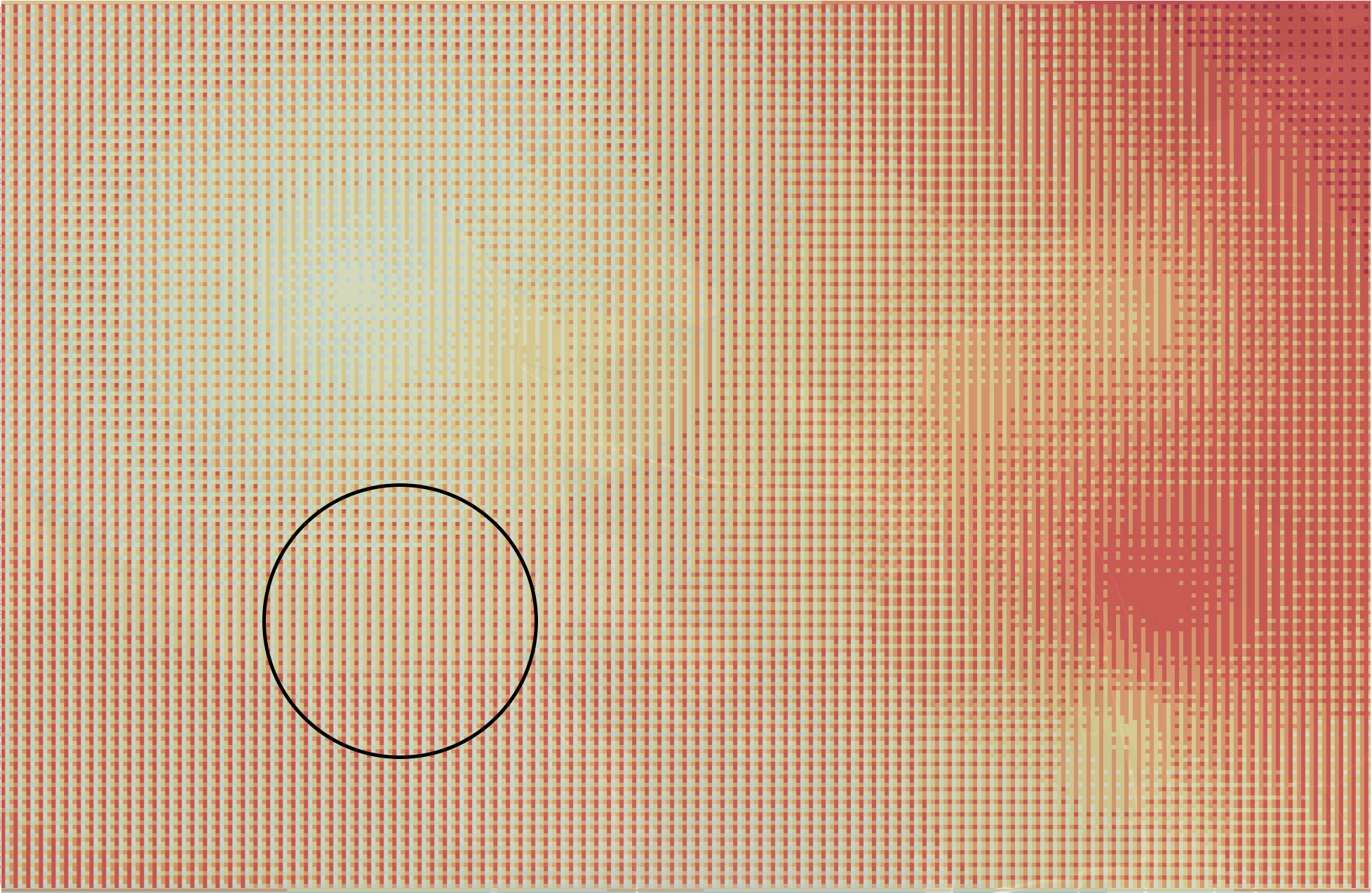}
        \caption{Ordered dotmap.}
        \label{dots_random}
    \end{subfigure}\hfill%
    \begin{subfigure}{0.45\textwidth}
        \centering
        \includegraphics[width=\textwidth]{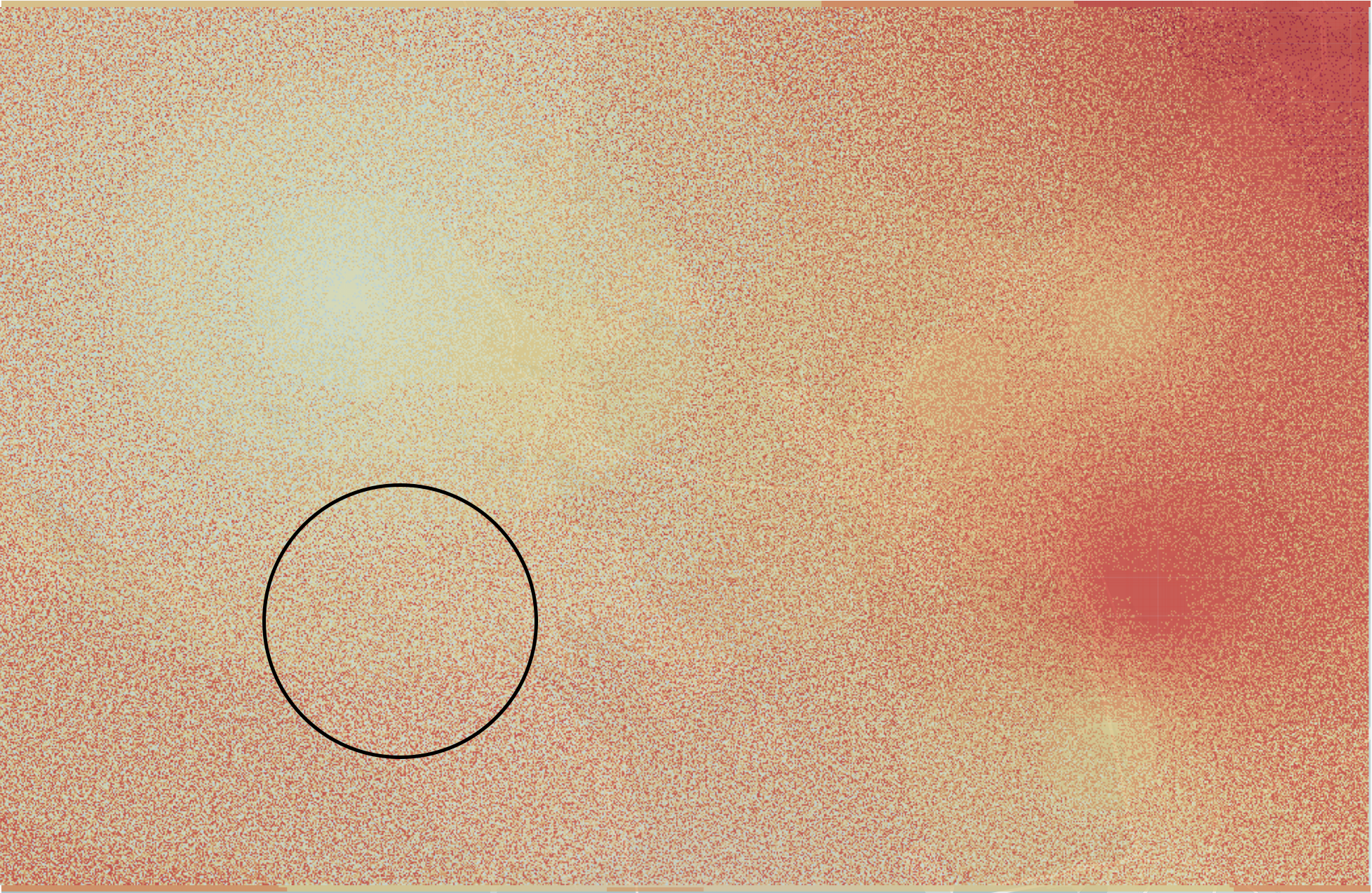}
        \caption{Smoothed dotmap.}
        \label{dots_ordered}
    \end{subfigure}\hfill%
   \begin{subfigure}{0.45\textwidth}
        \centering
        \includegraphics[width=\textwidth]{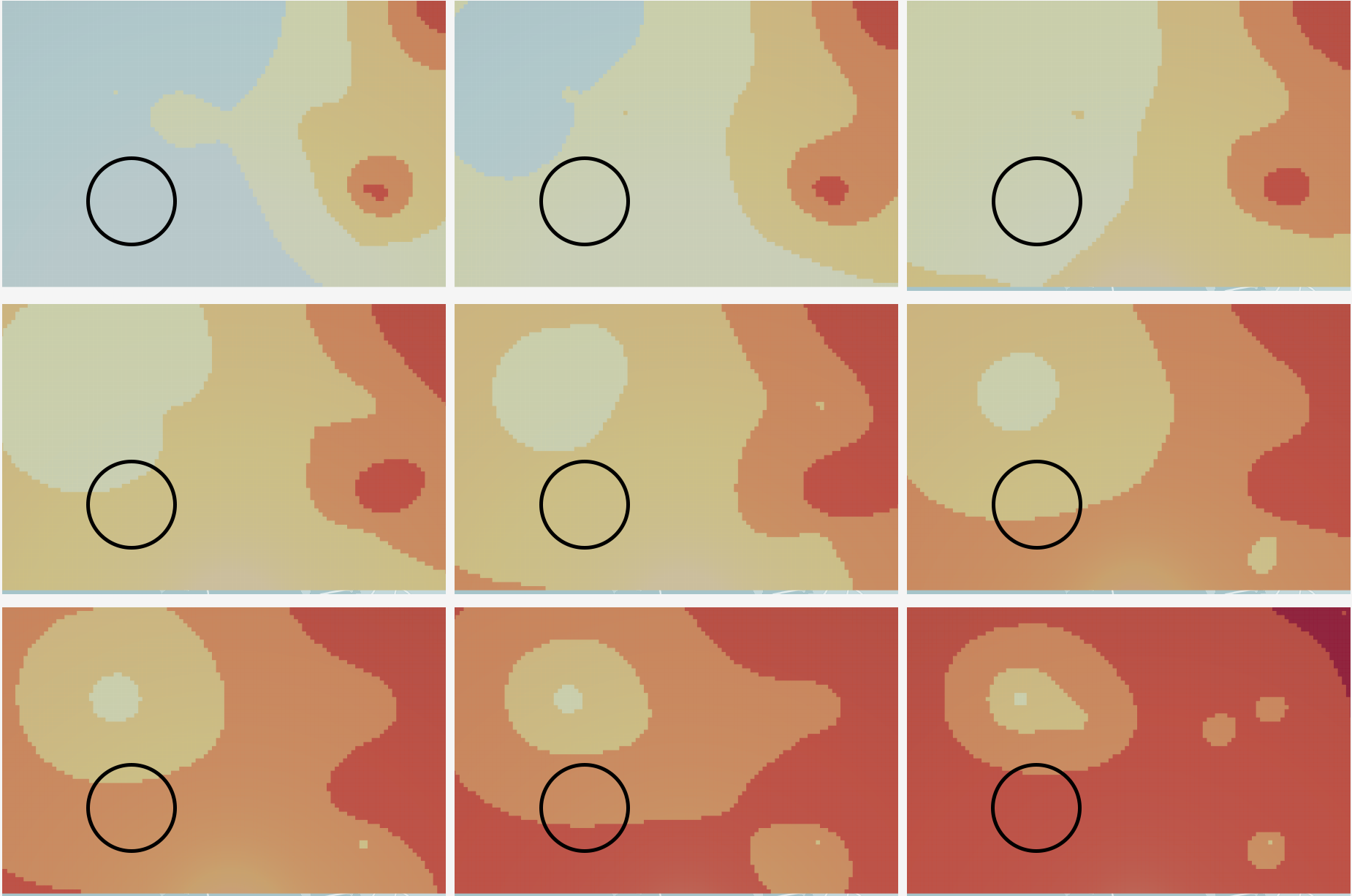}
        \caption{Small multiples.}
        \label{small_multiples}
    \end{subfigure}\hfill%
    \begin{subfigure}{0.45\textwidth}
        \centering
        \includegraphics[width=\textwidth]{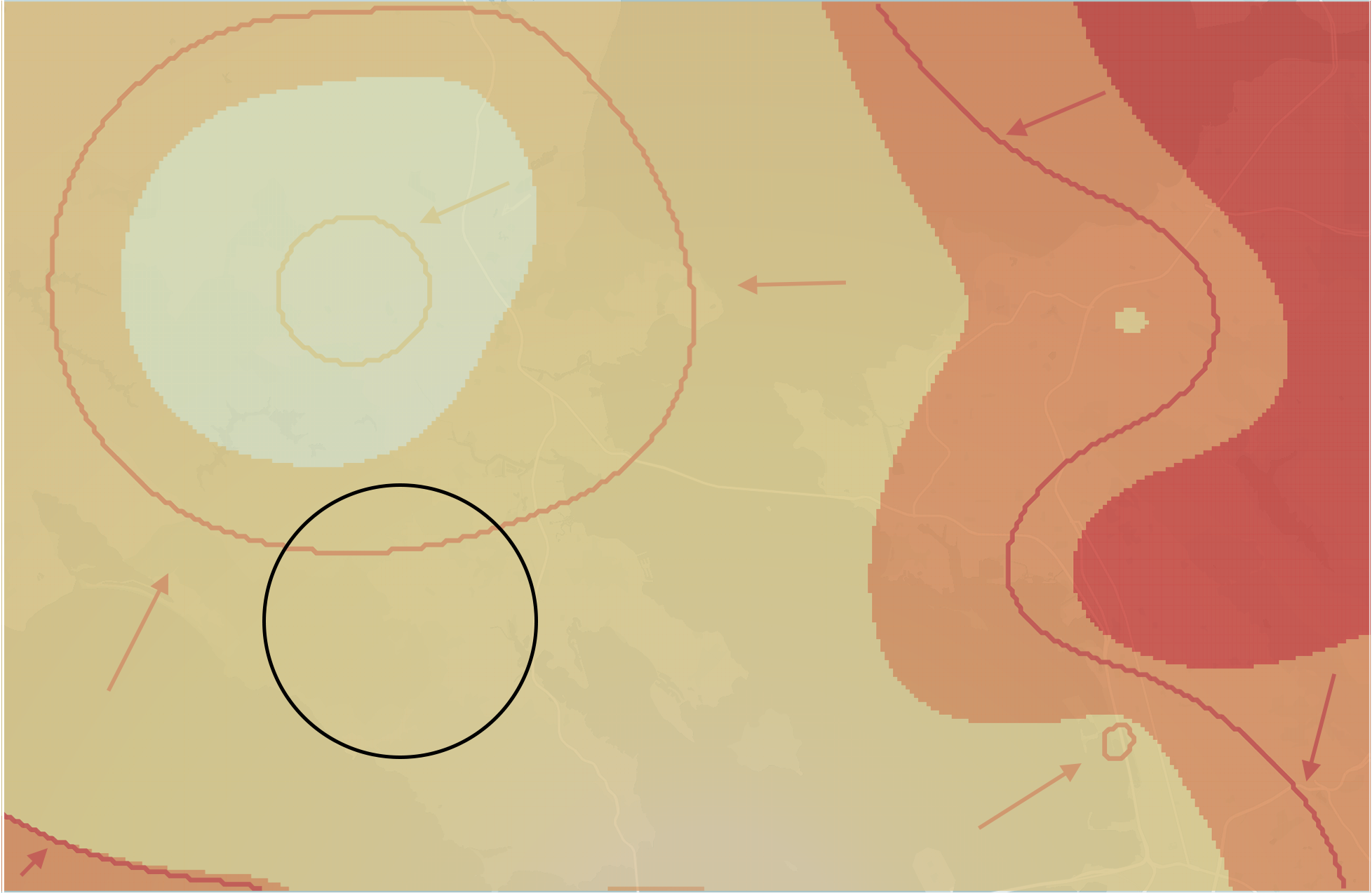}
        \caption{Risk contours.}
        \label{dots_synthesis}
    \end{subfigure}
     \caption{A set of stimuli for one of the user study scenarios. Users were asked to assume that they live and exercise primarily within the circle indicated on the map.}
     \label{stimuli_example}
\end{figure*}

\section{Evaluation}

Evaluating uncertainty visualizations is notoriously difficult; defining evaluation tasks that consider uncertainty is much more complex than those that do not~\cite{hullman2016}. In one similar set of studies evaluating uncertainty-aware visualizations of bus arrival times, users were asked to decide when to reach the bus stop in a given scenario, then they were shown the outcome (the bus's actual arrival time)~\cite{2016-when-ish-is-my-bus}~\cite{2018-uncertainty-bus}. Users tended to learn to use the uncertainty visualization designs to make better decisions over the trials, perhaps aided by seeing the outcome of their decisions immediately.

With air quality visualization, defining tasks is even more difficult. The decisions that one might  make in response to an air quality visualization are more categorical than numerical (e.g., choosing to take actions such as wearing a mask or staying indoors). There is no meaningful immediate feedback to give on a user's decision: the effects of air pollution are often hidden and long-term, and in real life, people have very little indication of how good their decision-making around air pollution is.

For these reasons, we focus on aspects of the decision-making process, rather than the outcomes of users' choices. One desired result is a set of explanations for how users reach decisions in each view. We consider relative changes in answers, and users' confidence in their answers, rather than soliciting probabilities directly, which may not translate well to real-world decisions. These suggestions are outlined in a recent survey of uncertainty visualization evaluation~\cite{2019-uncertainty-eval-survey}.

Our primary question is: \textit{How does the uncertainty representation affect people's decision-making?} Does their understanding of the data change with different uncertainty representations? We especially want to investigate how an explicit representation of the probability distribution compares with an implicit suggestion of uncertainty.

\subsection{User Study}

Applying our study to the real world assumes that people will put effort into understanding their air pollution risk, and would make the same effort in reality. Answers to this study may or may not reflect actions people would take in their lives. To mitigate this, we recruited members of our local community who had been exposed to fire-related air pollution within the past year, corresponding to the sample population in Section~\ref{aqa}. Personal experience has a significant impact on how people interpret maps, and this group has a relatively small range of personal experiences with air pollution compared with the global population, which may help us hone in on the factors that people use to make decisions with these maps. However, we must be cautious generalizing results to the general public. Of the 17 individuals (age 20-30) in the study group (7 female/10 male), 12 were Computer Science students and 5 were employed in other fields.

The in-person study followed a within-subjects design, where each participant ($n=17$) saw each of five scenarios in each of six map types (30 total stimuli). We use \textit{scenario} to mean a particular time and location, with all available air quality readings from PurpleAir and government sensors. Each scenario was generated within a bounding box near an urban region in northern California. We chose dates during 2017 and 2018 with significant air pollution from wildfires in these areas
(for example, see Figure~\ref{stimuli_example}). (Participants were only told that these were real scenarios.) Each stimulus image contained a circle representing the region within which the user would live and be active outside. Each scenario had one circle location that was used for all six map types. The circle locations were chosen to be away from edges and to contain some amount of uncertainty. Each participant used the same computer, which had a large enough screen size for the grid cells in the ordered dotmaps to be discernible.

We first performed a pilot study of two test subjects, using three of the scenarios. In the full study, each user was shown each stimulus, first the interpolation-only view for each scenario, then each other stimulus, in a rotating order so that scenarios were not repeated before seeing each of the others.  The order of scenarios and map types was balanced for each user, except for the interpolation-only map type. For example, if a user was assigned the scenario order $S_0$, $S_1$, $S_2$, $S_3$, $S_4$ and the map type order $M_0$, $M_1$, $M_2$, $M_3$, $M_4$, they would first see: $S_0$ $\times$ interp.-only, $S_1$ $\times$ interp.-only, ..., $S_4$ $\times$ interp.-only. Then, they would rotate through each combination: $S_0$ $\times$ $M_0$, $S_1$ $\times$ $M_1$, $S_2$ $\times$ $M_2$, $S_3$ $\times$ $M_3$, $S_4$ $\times$ $M_4$; then $S_0$ $\times$ $M_1$, $S_1$ $\times$ $M_2$, and so on.

Before the study, each user was guided through the same presentation which showed and explained each of the map types for a sample scenario; they were allowed to ask clarifying questions before starting the study.



For each stimulus, users were asked three questions:
\begin{itemize}
    \item \textbf{Q1.} If you had plans to run/bike outside today, would you reduce your plans? \textit{User answers on a 7-point scale from ``strongly disagree'' to ``strongly agree.''}
    \item \textbf{Q2.} Where, if anywhere, would you go for relief? \textit{User clicks on a point on the map.}
    \item \textbf{Q3.} How confident are you in your answer? \textit{User answers on a 5-point scale from ``not at all confident'' to ``strongly confident.''}
\end{itemize}

Users were asked to think aloud as much as possible while answering; we transcribed these answers and then identified different types of phrases from among the responses. Q2 was included to elicit more discussion from users about their decision-making process. Next to each stimulus, a legend appeared showing the colors corresponding to each AQI level (good, moderate, unhealthy for sensitive groups, unhealthy, very unhealthy, hazardous). Each of 17 users answered the questions using all six map types for all five scenarios, for a total of 510 observations.

\section{Results}
To assess whether map type influenced users' reported reduction in physical activity and their confidence in their decisions, we analyzed responses to Q1 ($P$) and Q3 ($C$) using generalized linear mixed models. We also performed non-parameteric tests for $P$ and $C$. 
\subsection{Quantitative Analysis: Physical Activity Change}
For $P$, the model included \textit{scenario} ($S$), \textit{map type} ($M$), and the \textit{interaction between scenario and map type} ($S \ast M$) as fixed effects; we included \textit{individual} ($I$) as a random effect to account for the fact that an individual's responses to different stimuli are not independent. The sample included 510 responses (17 users
$\times$ 5 scenarios $\times$ 6 map types).

$$ P \sim S + M + S \ast M  + 1 | I $$

We found that scenario ($df=4, F=113.5, p<0.0001$), map type ($df=5, F=12.1, p<0.0001$), and scenario $\times$ map type ($df=20, F=3.1, p=0.0001$)  all had highly significant effects on $P$. 

\begin{figure}[h]
\centering
        \includegraphics[width=0.48\textwidth]{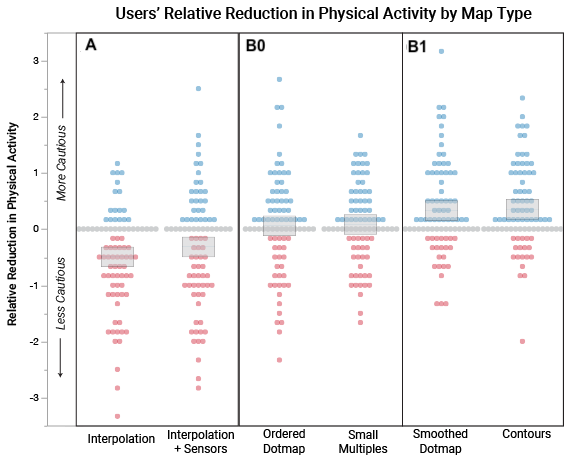}
        \caption{Values for users' answers to Q1, grouped by map type, after normalizing by individual and by scenario. Positive values (blue) reflect a higher reduction than the user's average for that scenario; negative values (red) reflect a lower reduction than average. Mean and standard error are within the gray boxes. The differences are statistically significant for map types between groups A, B$_0$, and B$_1$.}
        \label{activity_change}
\end{figure}

To differentiate the effects of map types---ones that include uncertainty versus ones that do not---we performed a post-hoc least-squares mean contrast, finding that $P$ was significantly lower for the Interpolation and Interpolation + Sensors views than for the other four views ($F=50.9, p<0.0001$). That is, including uncertainty in a view led to a higher reduction in physical activity. Within the uncertainty map types, the Ordered Dotmap and Small Multiples led to lower $P$ responses than the Smoothed Dotmap and Contours views ($F=7.54, p<0.0063$) (Table~\ref{table:1}).

Still, the effect of map type was highly dependent on scenario. Specifically, in one scenario, which showed extremely high air pollution levels, individuals chose to reduce their activity regardless of map type, leading to a significant effect of $S \ast M$. Prior work suggests that the risk level has a higher effect on users' responses than the visual features of the map, such as contours, focus, and how risk is encoded~\cite{Severtson2013}.

\begin{table}[t]
\centering
\begin{tabular}{ |c|c|c|c| } 
 \hline
 \textbf{map type} & \textbf{mean $P$} & \textbf{std. dev.} & \textbf{std. error} \\
 \hline
 interpolation & 4.76 & 1.58  & 0.17\\ 
 interpolation + sensors & 4.92 & 1.76 & 0.19\\ 
 \hline
 ordered dotmap & 5.32 & 1.43 & 0.16\\ 
 small multiples & 5.34 & 1.28 & 0.14 \\ 
 \hline
 smoothed dotmap & 5.59 & 1.22 & 0.13 \\ 
 contours & 5.61 & 1.01 & 0.11\\ 
 \hline
\end{tabular}
 \caption{Mean response to Q1. Each of the outlined pairs has significantly different effects on users' responses to Q1.}
\label{table:1}
\end{table}


 To check that this analysis is robust to deviations from normalcy, we used a matched pairs non-parametric Wilcoxon Signed Rank test. For a given individual in a given scenario, responses to Q1 ($P$) for the Interpolation and Interpolation + Sensors map types were, on average, significantly lower than responses for the other four views (Test Statistic: $-883, p<0.0001$). In 54/85 cases, mean participant $P$ response to the non-uncertainty map types was lower than it was for the four uncertainty map types, compared to only 23/85 times where the other pattern was observed (and 8/85 where they were equal).
 This test showed a less significant difference between $P$ responses for the Ordered Dotmap and Small Multiples views vs. Smoothed Dotmap and Contours views (Test Statistic: $-442, p=0.05$). 

These results suggest that our six visualization designs can be categorized into two, or perhaps three, types, each with different effects on users' decisions (see Figure~\ref{activity_change}). The first type (group A) includes the standard map types: interpolation only and interpolation with sensors. Users were most willing to continue their exercise outside when judging air quality based on these maps. The second group includes the uncertainty views, which may be broken into two categories. Group $B_0$ includes the frequency-framing designs: small multiples and the ordered dotmap. (Though smoothed dotmaps are created with the same frequency information as the maps in group $B_0$, their resolution makes it difficult to discern frequency.) Users' responses to the map types in group $B_0$ were intermediate. The maps in group $B_1$ (risk contours and smoothed dotmap) are the non-frequency-framing uncertainty views. Users reduced physical activity the most in response to these maps.

The consistency of responses for $P$ also varied depending on map type. We used a loglinear variance modeling approach and found that in a model including scenario and map type as mean effects, and map type as a variance effect, map type had a significant effect on the variability of a participant's $P$ responses ($df=5, \chi^2=29.4, p<0.0001$). Variability of responses was highest for the Interpolation + Sensors view, followed by Interpolation only. (In section~\ref{thinkaloud}, we discuss reasons this might be the case.)

\begin{figure}[h]
\centering
        \includegraphics[width=0.48\textwidth]{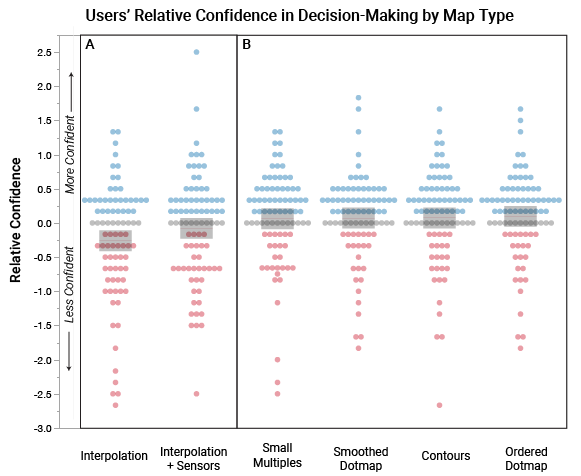}
        \caption{Q3 answers, grouped by map type, after normalizing for scenario and individual. Positive values (blue) reflect higher confidence relative to the user's average for that scenario; negative values (red), lower confidence. Mean and standard error are within the gray boxes. The difference in users' responses is statistically significant between map types in groups A and B.}
        \label{confidence_change}
\end{figure}
\subsection{Quantitative Analysis: Confidence}
We created a similar generalized linear mixed model to analyze users' confidence in their responses ($C$):
$$ C \sim S + M + S \ast M + 1 | I$$

While scenario still had a highly significant effect on $C$ ($F=5.3, p=0.0004$), map type showed a less significant effect ($F=2.3, p=0.041$), and scenario $\ast$ map type was not significant ($F=0.9, p=0.61$). Using a Least Squares Mean Contrast, we found that the Interpolation and Interpolation + Sensors views led to significantly lower confidence than the other views (F=9.8, p=0.002), but there was no significant difference between Ordered Dotmaps and Small Multiples vs. Smoothed Dotmaps and Contours ($F=0.005, p=0.95$). Wilcoxon Signed Rank tests confirmed a significant difference between Interpolation and Interpolation + Sensors versus the uncertainty map types (Test Statistic$=-532, p=0.016$).

\begin{table}[t]
\centering
\begin{tabular}{ |c|c|c|c| } 
 \hline
 \textbf{map type} & \textbf{mean $C$} & {std. dev.} & {std. error}\\
 \hline
 interpolation + sensors & 3.73 & 1.09 & 0.12\\ 
  interpolation & 3.89 & 1.04 & 0.11\\
  \hline
small multiples & 4.04 & 0.82 & 0.09 \\
smoothed dotmap & 4.04 & 0.97 & 0.11 \\
contours & 4.05 & 1.00 & 0.11 \\
ordered dotmap & 4.07 & 1.00 & 0.11 \\
 \hline
\end{tabular}
 \caption{Mean response to Q3. The first two map types have a significantly different effect on users' confidence than the other four map types.}
\label{table:2}
\end{table}


Users' confidence was significantly higher for the uncertainty views than for the two standard views (see Figure~\ref{confidence_change}). One caveat is that because we are not considering the ``correctness'' of users' answers, we cannot correct for the ``hard-easy effect'' of confidence reporting~\cite{2019-uncertainty-eval-survey}. Users often deliberated for longer while making their decisions for the uncertainty views, perhaps resulting in users perceiving these decisions as more difficult and therefore reporting higher confidence ratings for these map types. 

\begin{figure*}[h!]
\centering
\includegraphics[width=\linewidth]{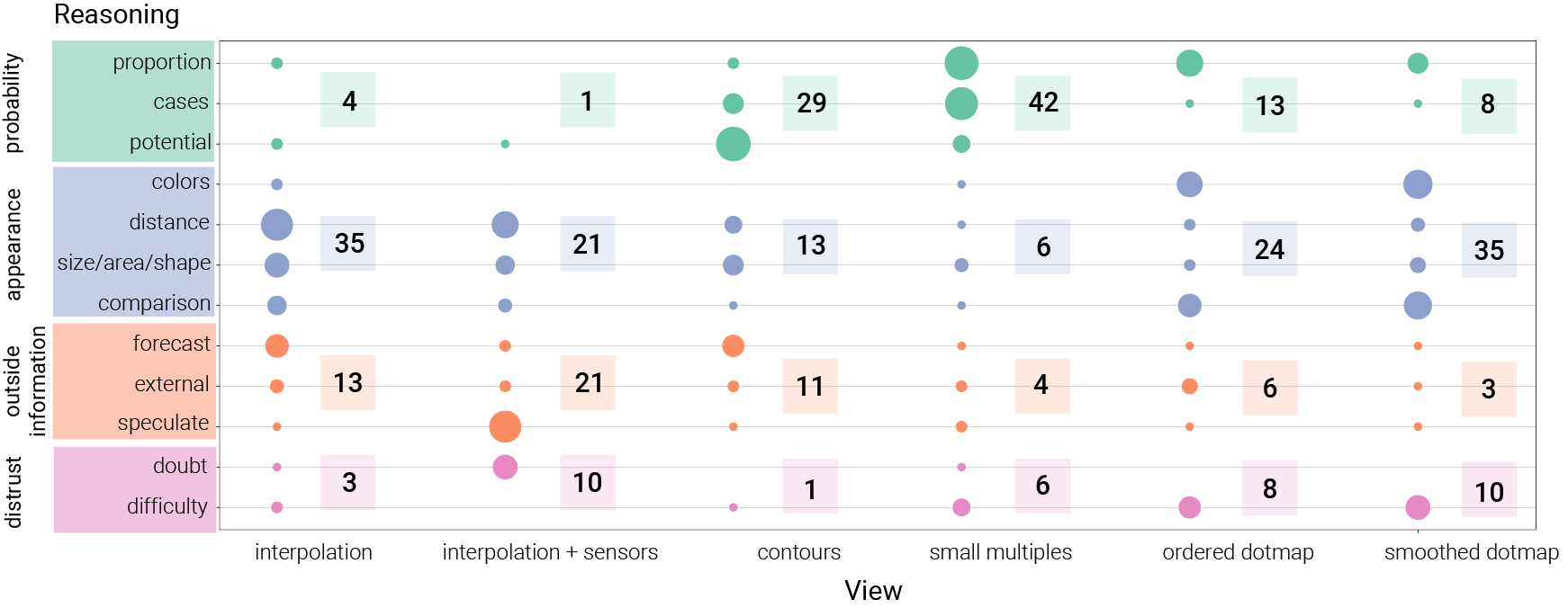}
\caption{Occurrences of each type of reasoning that was spoken aloud for each view. Outside information use was highest for the non-uncertainty views, especially \textit{interpolation + sensors}, which also led users to express the most doubt. Frequency-based reasoning, especially using proportions or best and worse case scenarios, was very common for small multiples. Both types of dotmaps led users to express similar types of reasoning.}
\label{comments_views}
\end{figure*}

\subsection{Think-Aloud Results}
\label{thinkaloud}

We analyzed think-aloud transcriptions to understand how users made decisions with each of the map types.
We transcribed users' think-aloud feedback during the studies, then identified types of phrases or reasoning that came up repeatedly, grouping them into four categories:

\noindent \textbf{Probability}
\begin{itemize}
    \item \textbf{proportion}: the user discusses ratios, such as ``six of the nine say it's unhealthy'' or ``half and half.''
    \item \textbf{cases}: the user considers the best, worst, or likeliest scenarios, usually using that phrasing.
    \item \textbf{potential}: the user talks about what the pollution level ``could'' or ``might'' be.
\end{itemize}

\noindent \textbf{Appearance}
\begin{itemize}
    \item \textbf{colors}: the user describes which colors are present in the map, and their amounts.
    \item \textbf{distance}: the user makes judgments based on relative distances away from areas of concentrated pollution.
    \item \textbf{shape, size, area}: the user makes judgments based on the shape of the contours, or the overall map area covered by a color.
    \item \textbf{comparison}: the user compares areas in the map to one another based on appearance.
\end{itemize}
\textbf{Outside Information}
\begin{itemize}
        \item \textbf{forecast}: the user speculates about how the pollution will behave over time.
        \item \textbf{external factors}: the user discusses their own potential factors such as a friend coming into town, or says that they would see how they feel on that day.
        \item \textbf{speculate}: the user speculates about information not specified in the map, like nearby air polllution.
\end{itemize}
\textbf{Distrust}
    \begin{itemize}
        \item \textbf{doubt}: the user disbelieves the data shown or expresses that an optimal choice cannot be made.
        \item \textbf{difficulty}: the user expresses difficulty in making a choice or interpreting a map.
    \end{itemize}

Figure~\ref{comments_views} shows the number of occurrences of each of these categories per view. Several patterns emerge:


\subsubsection{Non-Uncertainty Views Encourage Qualitative Judgments and Outside Information}
For views without any uncertainty information, users most often used external information to supplement the maps. With the interpolation-only view, users considered shape, size, area, or distance, often in conjunction with speculating about variation over time or personal factors. With the added information of the sensor data, users relied on these factors less, often identifying aloud the colors in the map. However, users in the interpolation-and-sensors case more often used broad qualitative judgments and expressed skepticism about the information. Decision-making became dependent on personal judgment. Those with more experience with interpolating sensor data preferred this: \textit{``Showing the sensors is good because I can build up my own interpolations.''}

However, many users expressed doubt with the interpolation+sensors view, and, as shown in the quantitative analysis, these views resulted in the widest variance in user response. Research suggests that most people dislike ambiguity, preferring more certain information about risks~\cite{Severtson2013}. This research finds that this ambiguity may increase or decrease people's risk reduction related to health, and that personal experience also has a strong effect on people's interpretation of visualizations and maps. For a broader study population reflective of the general public, these views are likely to result in a wide range of interpretations, varying due to individuals' background knowledge and personal response to ambiguous risk information. Designs that encourage reasoning independent of personal experience are more likely to translate successfully to the general public.

\subsubsection{Small Multiples Encourage Frequency Reasoning}


Users mentioned probabilities and ratios most frequently for the small multiples view. Some users were initially confused by this view, but many developed their own strategies for interpreting it over the course of the study, and some commented in particular on its utility. 

\textit{``I feel that it's easier than I thought to use the nine views because I can compare, and I can see the worst case.  If it's the worst case, I won't go there, it made me confident.''}

The results from the quantitative analysis---small multiples yielded the least varied responses (see Figure~\ref{activity_change})---suggest that in general, users were able to apply consistent reasoning to the small multiple maps. However, users occasionally had low confidence when using the small multiples; many people may need more help interpreting these maps.

\subsubsection{Dotmaps: Similar Reasoning at Both Resolutions}

The reasoning types were similar between the ordered dotmap and smoothed dotmap, including a similar amount of difficulty. The quantitative results suggest that though reasoning was similar, the ordered dotmap led to more cautious answers than the smoothed dotmap. Ordered dotmaps and small multiples led to similar decisions, but frequency-based reasoning was vocalized more for small multiples.

Particular visual features may contribute to increased risk perception in the smoothed dotmaps. These maps often left users uncertain of which colors in the AQI legend were being represented, but gave them the impression of ``a lot of bad dots'' and, therefore, increased risk. This effect indicates that users' interpretation of the smoothed dotmap view was more similar to ``noise annotation lines''~\cite{kinkeldey_2014}.

The dotmap views were often the most difficult to interpret, particularly the smoothed version. This was most true with greater variation, since it became difficult to pick out individual colors. Some of the stimuli lended themselves to easier judgments about ratios for ordered dotmaps; users mentioned proportion the second-most frequently using this view. Users also again turned to qualitative judgments or impressions of the shapes of polluted areas, sometimes commenting that these maps made it easier to spot patterns than to identify individual AQI values. 

One user preferred these views, saying that the ordered dotmaps are more effective when there is less variation:

\textit{``I like the one like random dots, and the little squares, the matrix ones - those are relatively the same to me...but I prefer the random one, because it's more smoothly spread. Except for one case, there is an area that's all yellow but others, there are lines and dots; for that one I notice it's different and it's easier to use the matrix one. Maybe for some cases, this one is better and for some cases, that one is better.''}

Users' ease in interpreting the dotmap views was  scenario-dependent. More work is needed to figure out optimal grid sizes given an amount of variation in air quality.

\subsubsection{Contour Maps Suggest High Risk Potential}

Looking at the contour view, users were most likely to express \textit{potential}, such as, ``this area could be orange.'' Identifying this potential often corresponded with choosing a more cautious answer for the stimulus. This was sometimes misinterpreted as showing how the air pollution might evolve over time---an example of a deterministic construal error, in which an easier explanation is substituted for a more difficult one~\cite{doi:10.1177/0963721413481473}---but decisions based on ``potential'' and ``forecast'' were similar despite different rationales.  The contour view also often yielded the fastest and most decisive-seeming answers. Some users mentioned feeling particularly confident with the risk contour view: 

\textit{``I think the arrows did a good job of me being confident in a region being fine when you could see some boundary that...ended.''}

\textit{``If I need to spend more time on analyzing the visualization, I tend to have more confidence. So the boundaries with the arrows, I feel more confident about those visualizations.''}

Prior work supports the idea that users prefer the ``certainty'' of a contoured map like this, even though it is a simplification of the underlying range of risk levels. Compared to unfocused views like the dotmaps, focused contours may result in stronger beliefs for higher risk levels~\cite{Severtson2013}.

\section{Conclusion}

Our results support that uncertainty information can help users make decisions more confidently and with a higher perception of risk, and that the choice of visualization significantly affects users' decisions. 
In particular, including uncertainty information made people more cautious.
In line with recent research on uncertainty visualization and decision-making in other domains, our results suggest that users were most able to optimize their decisions---align their choice with their risk tolerance---using a frequency-framing rationale with a small multiples view.
Standard maps that show no or implicit uncertainty result in a more unpredictable range of user responses, while using discretized uncertainty may encourage more consistent responses, allowing users to apply robust reasoning. 

There are some important limitations to these findings that we hope can be addressed in future studies. First, our sample population was chosen for its specific recent experience with wildfire smoke, but generalizing these maps to a wider population will involve a much broader range of personal experience. While the users in our study were able to ask clarifying questions about the maps, we did not explicitly test to see whether users understand how to correctly interpret the map types, just as might happen in the real world. More work is needed to ensure that people can interpret uncertainty views correctly, especially outside of a study environment. Finally, to mimic a typical experience, we showed the interpolation-only views to users first, before they saw other views, so that their reasoning would be closer to reasoning in the real world. Our findings on non-uncertainty views compared to the others included the interpolation+sensors map, which was shown in the blocking order among the others. Still, there may have been a bias resulting from this ordering choice.

Our results show potential for people to use uncertainty-based maps to understand environmental risks, but the designs presented here can be improved upon by combining some of their strengths and optimizing features. We found that scenario had a strong effect on users' reported reduction in physical activity, meaning that scenario and map type both determine interpretation; more work is needed to understand how each design behaves with a real range of datasets. A larger follow-up study on a broader population could inform designs that can be adopted by the general public under a range of scenarios. 

\section*{Acknowledgments}
The authors thank Sandra Bae (UC Davis) for discussions on study design; Prof. Anthony Wexler (UC Davis) for advice on uncertainty in air quality sensors; and Dr. Jack Colicchio (UC Berkeley) for feedback on the quantitative analysis. This research was sponsored in part by the U.S. National Science Foundation through grants IIS-1741536 and IIS-1528203. 


\ifCLASSOPTIONcaptionsoff
  \newpage
\fi



\bibliographystyle{IEEEtran}
\bibliography{IEEEabrv,main}
\end{document}